\documentclass[11pt,epsf,epsfig,psfrag]{article}
\usepackage{lineno}
\usepackage{hyperref}
\usepackage[bf]{caption}
\usepackage{epsfig}
\usepackage{psfrag}
\usepackage{color}
\usepackage{amsmath}
\usepackage[mathscr]{euscript}
\usepackage[toc,page]{appendix}

\usepackage{titlesec}
\titleformat{\paragraph}
{\normalfont\normalsize\bfseries}{\theparagraph}{1em}{}
\titlespacing*{\paragraph}
{0pt}{3.25ex plus 1ex minus .2ex}{1.5ex plus .2ex}

\usepackage{float}
\usepackage{graphicx}
\usepackage{epstopdf}
\usepackage{subfig}
\def\ga{\mathrel{\raise.3ex\hbox{$>$\kern-.75em\lower1ex\hbox{$\sim$}}}}
\def\la{\mathrel{\raise.3ex\hbox{$<$\kern-.75em\lower1ex\hbox{$\sim$}}}}

\def\I_M{{I_{\scriptscriptstyle M\times M}}}

\def\lsim{\mathrel{\rlap{\lower4pt\hbox{\hskip1pt$\sim$}}
    \raise1pt\hbox{$<$}}}                
\def\gsim{\mathrel{\rlap{\lower4pt\hbox{\hskip1pt$\sim$}}
    \raise1pt\hbox{$>$}}}   

\begin{document}

\thispagestyle{empty}

\vskip 2cm

\begin{center}{\Large \bf Thermodynamic geometry of the spin-1 model. II. Criticality and coexistence in the mean field approximation}
\end{center}

\vskip .2cm

\vskip 1.2cm

\centerline{ \bf Anurag Sahay, Riekshika Sanwari\footnote{anuragsahay@nitp.ac.in, riekshikas.phd19.ph@nitp.ac.in  } 
}

\vskip 4mm\centerline{ \it  Department of Physics, National Institute of Technology, Patna 800005,  India}

\vskip 1.2cm
\vskip 1.2cm
\centerline{\bf Abstract}
\noindent

We continue our study of the thermodynamic geometry of the spin one model from \cite{reiksh} (paper I) by probing the state space geometry of the Blume Emery Griffiths  (BEG) model, and its limiting case of the Blume Capel model, in their mean field approximation. By accounting for the stochastic variables involved we construct from the thermodynamic state space two complimentary two-dimensional geometries with curvatures $R_m$ and $R_q$ which are shown to encode correlations in the model's two order parameters, namely, the magnetization $m$ and the quadrupole moment $q$. The geometry is investigated in the zero as well as the non zero magnetic field region. We find that the relevant scalar curvatures diverge to negative infinity along the critical lines with the correct scaling and amplitude. We then probe the geometry of phase coexistence and find that the relevant curvatures predict the coexistence curve remarkably well via their respective $R$-crossing diagrams. We also briefly comment on the effectiveness of the geometric correlation length compared to the commonly used Ornstein-Zernicke type correlation length vis-a-vis their scaling properties.
\setcounter{footnote}{0}
\noindent

\section{Introduction}
\label{intro}

In \cite{reiksh} (paper I) we had established the thermodynamic geometry of the one dimensional spin one model starting from its exact free energy obtained via the transfer matrix calculation. Along with the full three dimensional scalar curvature two sectional curvatures $R_m$ and $R_q$ were also worked out which were found to correctly encode the correlation lengths for spin-spin and quadrupole-quadrupole correlations respectively. We also extensively verified the `strong' Ruppeiner conjecture relating the inverse of the critical free energy to the scalar curvature,
\begin{equation}
R=\frac{\kappa}{\psi}
\label{Ruppeiner eq}
\end{equation}
which we refer to as the `Ruppeiner equation'. Remarkably, the association of the scalar curvatures with the calculated correlation lengths often extended to regions far beyond criticality, thus corroborating what we termed the weak Ruppeiner conjecture. 

In this work we extend our geometrical analysis to the mean field spin one model as obtained in \cite{beg}. The mean field case immediately poses a challenge since, unlike the exactly solvable one dimensional case, there is no calculated correlation length to compare the scalar curvature with. At the same time, looked the other way round, it is also an opportunity since, unlike the one dimensional case where our job was essentially that of verification against an already calculated quantity, the mean field model in itself has no scope for a correlation length. Allowing for a slow, predetermined variation of the order parameter results in a Landau-Ginzburg free energy from which one obtains an expression for the correlation length, \cite{pathria}. On the other hand, the Riemannian geometry arising out of the mean field equations of state has an intrinsic curvature scale which is conjectured to be the correlation volume of the underlying system, \cite{rupprev}. The geometrical construction produces a scalar curvature as a state function characterizing the volume scale of the physical system below which the assumptions of classical fluctuation theory break down, arguably because the fluctuations below this volume scale are coherent. In other words it is the correlation volume. Geometry therefore ends up enriching the information content of the mean field thermodynamics. In any case this is the claim we put to test here in the context of the mean field BEG model. Satisfyingly, our subsequent analysis substantiates this claim from several perspectives and also shines a light on the geometric correlation length as a somewhat better suited candidate to analysing the scaling behaviour in comparison to the commonly used Ornstein-Zernike like correlation length. Also, it is more satisfying, if somewhat mystifying, that we can obtain the correlation length from the equations of state itself without needing to extend the thermodynamics to include space dependent terms. 

The outline of the work is as follows. In section \ref{ha} we review the mean field model and its phase structure and in subsection \ref{haha} we carefully construct two thermodynamic geometries of the mean field model suited to correlations in its two order parameters. We then discuss the geometry near criticality in section \ref{hu} and near coexistence in section \ref{huha}. In subsection \ref{hai} we briefly discuss the scaling properties of the geometric correlation length as compared to the square gradient one. In subsections \ref{hai1}, \ref{hai2} and \ref{hai3} we discuss in turn the geometry near the zero field critical line, the tricritical point and finally near the wing critical line. In subsections
 \ref{fa} and \ref{fu} we try to construct from geometry the phase coexistence curves in zero field and in non-zero field regions respectively. Finally, in section \ref{ga} we conclude our work.

\section{The mean field BEG model, its phase structure and geometry}
\label{ha}

The BEG model has the most general reflection symmetric Hamiltonian for a classical spin one model with nearest neighbour interactions. The Hamiltonian for the BEG model is written as, \cite{beg}
\begin{equation}
 \mathcal{ H}_{\small{beg}} = -J\, \sum_{<ij>} \, S_i\, S_{j} - K\,\sum_{<ij>}\, S_i^{\,2}\, S_{j}^{\,2}-H\, \sum_{i} \, S_i+ D\,\sum_{i} S_i^{\,2}
\label{BEG}
\end{equation}

The lattice spin variable $S_i $ is Ising like and can take up values $+1,-1$ and $0$. In addition to the bilinear coupling terms and a magnetic field that couples to the magnetic moment, the Hamiltonian contains a biquadratic coupling term of strength $K$ and a crystal field $D$ which couples to the quadrupole moment. The coupling strengths $J$ and $K$ are positive in the original BEG model, \cite{beg}. The $K=0$ limit is the Blume Capel (BC) model, \cite{blume,capel}. While the magnetic field term is not experimentally realizable in the original context of the BEG model which refers to a mixture of  $\mbox{He}^3$-$\mbox{He}^4$, it plays its usual role in the BC limit which refers to a magnetic system. The spin one models have two densities\footnote{in fact these are defined as magnetization, etc per lattice, but since the number of lattice points scales as the volume we shall often use the terms interchangeably.}, namely the mean magnetization and the mean quadrupole moment,
\begin{equation}
 m=\langle S_i\rangle\,\,\,\,;\,\,\,\, q=\langle S_i^2\rangle
\label{order}
\end{equation}
 In the lattice gas interpretation of the BEG model $m$ represents the superfluid order parameter while $q$ is the concentration of $\mbox{He}^4$. In a magnetic system $x=1-q$ would measure the concentration of non magnetic impurities. Similarly, apart from a small term, $D$ is the difference in chemical potentials of $\mbox{He}^3$ and $\mbox{He}^4$, namely $D \sim \mu_3-\mu_4$. For a positive $D$ larger concentrations of $\mbox{He}^3$ would be energetically preferred. In the context of a magnetic system, $D$ refers to a single ion anisotropy term  which splits the single-spin energy levels, with $S_i =0$ lower than the degenerate $S_i=\pm1$ levels. It can also be thought of as an external field coupled to the order parameter $Q$ (or $x$) analogous to $H$ which couples to $M$. Owing to the interplay of the two order parameters via the interaction terms and the field terms the phase structure of the BEG and related models is rich, with the presence of tricritical point, the triple line, critical lines and a line of first order transitions, \cite{beg,blume,capel}. 
 
 The qualitative features of the phase structure are already captured in the mean field approximation to the model, to which we turn now. The mean field Hamiltonian can be written as
 \begin{equation}
  \mathcal{ H}_{\,\small{\,beg}}^{\small{mf}}=-(\frac{J z m}{2}+H)\sum_{i} \, S_i-(\frac{K z q}{2}-D)\sum_{i} \, S_i^2,
  \label{meanfieldbegeq}
 \end{equation}
 with $z$ being the co-ordination number of the lattice. In the sequel we shall scale all the quantities by the factor $J\,z$ but shall continue to represent the scaled quantities by their unscaled symbols, so that we have
 $ \mathcal{ H}_{beg}^{mf}/Jz\to\mathcal{H}_{beg}^{mf}$, $ H/Jz\to H$, $D/Jz\to D$, and for later reference, $ \beta Jz \to \beta$. The ratio $K/J$ will appear subsequently as $\omega$.
 
In the mean field Hamiltonian the effects of spin-spin and quadrupole-quadrupole interactions are approximated, respectively, by the effective $H$ field and the  effective $D$ field. This results in a single site Hamiltonian which is solvable, with the trade-off being that the magnetization $m$ and the quadrupole moment $q$ are to be obtained self consistently by a minimization of the mean field free energy which is obtainable from the partition function,

The self consistent expressions for the magnetic moment and the quadrupole moment are obtained as

\begin{equation}
m=\frac{2  \sinh \beta(m+H)}{2  \cosh \beta( m+H )+e^{\beta\,  D-\beta\,  q\, \omega }}
\label{magnetic moment}
\end{equation}
and
\begin{equation}
q=\frac{2  \cosh \beta(m+H)}{2  \cosh \beta( m+H )+e^{\beta\,  D-\beta\,  q\, \omega }}
\label{quad moment}
\end{equation}

Making use of the above expressions the logarithm of the partition function can be written as
\begin{equation}
\psi=\log \mathcal{Z}_{mf}=\log \frac{q}{1-q}-\frac{1}{2} \beta  \left(m^2+q^2\,\omega \right)
\label{massieu}
\end{equation}

and the free energy is
\begin{equation}
G=\frac{-\psi}{\beta}
\label{free energy}
\end{equation}
Of course, the $m$ and $q$ in eq.(\ref{massieu}) still need to satisfy the self consistent equations for equilibrium. Keeping in mind that in the BEG model it is the small values of $\omega$ that are physically relevant we shall limit our investigations to such cases here. Nevertheless, we mention that the cases of larger $\omega$ ($>1$) have an interesting phase structure, some details of which may be found in \cite{beg}.

In the superfluid phase where $m\neq0$ the quadrupole moment $q$ can be expressed in terms of $m$ as
\begin{equation}
q=m\,\coth \beta(m+H)
\label{q superfluid}
\end{equation}
and substituting this value for $q$ in eq.(\ref{magnetic moment}) above, $m$ can be found self consistently.
In the normal phase, $m=0$, $q$ can be found from its transcendental equation. Eqs.(\ref{magnetic moment})
and (\ref{quad moment}) simplify considerably in the Blume-Capel limit $\omega=0$.

\begin{figure}[t!]
\centering
\includegraphics[width=3.5in,height=2.7in]{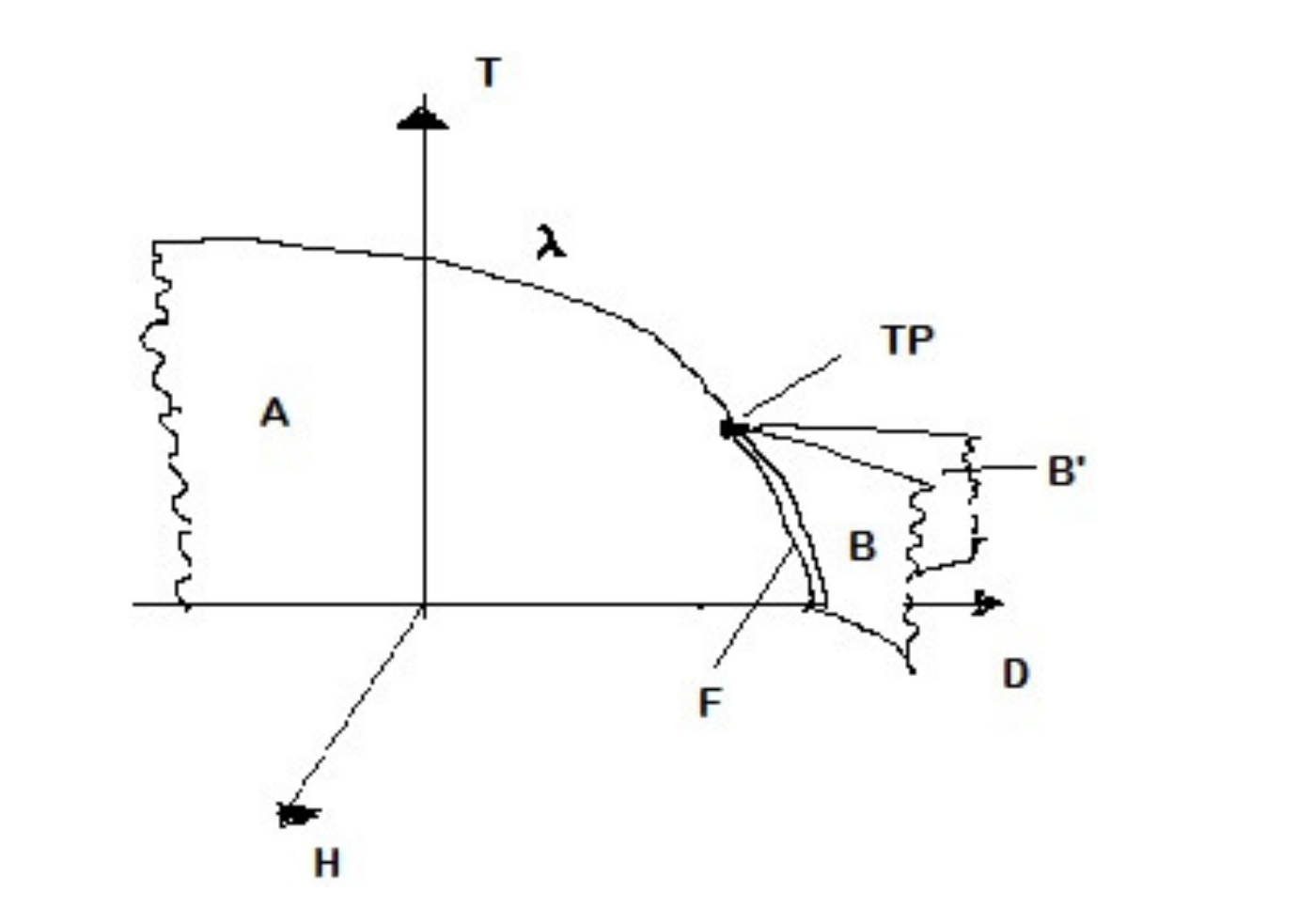}
\caption{Schematic diagram of the mean field phase structure of the spin one model in the $T-D-H$ plane. Here the ratio $K/J$ is small. Adapted from \cite{krinsky}}
\label{meanfieldbeg}
\end{figure}

In fig.\ref{meanfieldbeg} we present a mean field phase diagram of the Blume-Capel model ($\omega=0$ with the Hamiltonian given by eq.(\ref{BEG}). The qualitative features remian the same for small values of $\omega$, \cite{beg,mukamel}. Here  ${\bf A}$ is a coexistence surface in the $T-D$ plane where phases with positive and negative $M$ coexist. In the limit $D\to-\infty$ only the $S=-1,1$ states appear and the system is mapped to a spin half Ising model. In the $\mbox{He}^3$-$\mbox{He}^4$ context it would mean the absence of any $\mbox{He}^3$ impurity. Also, there are symmetrically placed wing like coexistence surfaces ${\bf B}$ and ${\bf B'}$ extending into the $D-H$ plane for $D>D_{tcr}$. The wings act as coexistence surfaces for the paramagnetic states having different values of $M$ and $Q$. The coexistence surface ${\bf A}$ and the wings each is bordered on the high temperature side with critical lines which join together at the tricritical point {\bf TP}. The line of intersection of the three coexistence surfaces in the triple line ${\bf F}$ which terminates at teh tricritical point. The exponents $\beta$, $\delta$ and $\alpha$ at the tricritical point are different from their critical values while the $\gamma$ and $\nu$ are the same, \cite{chaikin}

\begin{figure}[t!]
\begin{minipage}[b]{0.3\linewidth}
\centering
\includegraphics[width=2.3in,height=1.8in]{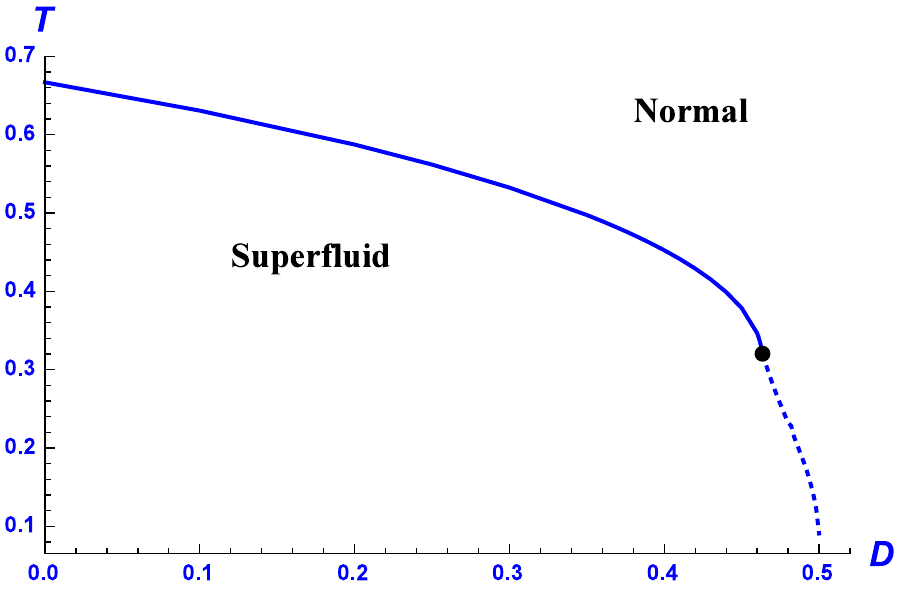}
\end{minipage}
\hspace{3cm}
\begin{minipage}[b]{0.3\linewidth}
\centering
\includegraphics[width=2.3in,height=1.8in]{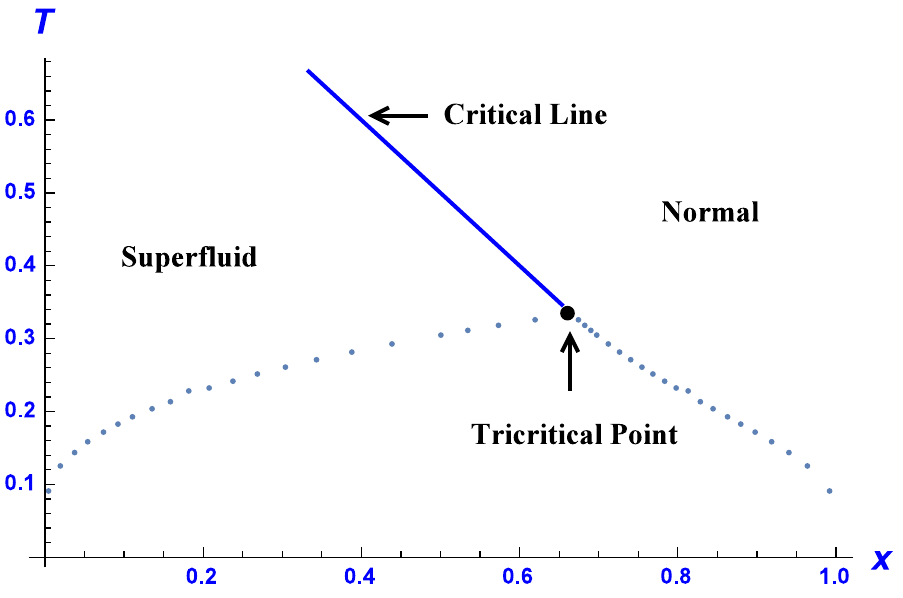}
\end{minipage}
\caption{\small{ $(a)$ Plot of the zero field critical and coexistence lines in the $D-T$ plane for the Blume-Capel model with $\omega=0$. $(b)$ The black dot is the tricritical point at $D=0.4621$ and $T=1/3$, while the dotted line below is the coexistence curve and the smooth line above is the critical line. $(b)$ The same phase structure in the $x-T$ plane where $x=1-q$. The coexistence region is covered by the dotted curves. Both sub-figures are adapted from \cite{beg}. }}
\label{phase1}
\end{figure}

In fig. \ref{phase1}(a) we plot the zero field critical line, tricritical point and the coexistence line in the $D-T$ plane for the Blume-Capel limit.  As we have mentioned earlier, the phase structure remains the same for small values of $\omega$ in the BEG mdoel. Referring the readers to \cite{beg,strecka} for detailed derivations, we briefly outline here some mathematical relations pertaining to the phase structure. Firstly, in the Blume-Capel limit the zero field expression for magnetization simplifies considerably,
\begin{equation}
m=\frac{2  \sinh (\beta  m)}{2  \cosh (\beta  m)+e^{\beta\,D}}\hspace{2cm}(\mbox{Blume-Capel},\,H=0)
\label{mblumecapel}
\end{equation}
and in the vicinity of the critical point the Landau expansion of the free energy gives
\begin{equation}
G=-\frac{\log \left(2 e^{\beta  (-D)}+1\right)}{\beta }+ \left(\frac{1}{2}-\frac{\beta }{e^{\beta  D}+2}\right)\,m^2-\frac{ \left(\beta ^3 \left(e^{\beta  D}-4\right)\right)}{12 \left(e^{\beta  D}+2\right)^2}\,m^4+...
\label{Landau}
\end{equation}
The critical points are obtained via the condition $\partial^2 G/\partial m^2=0$ at $m=0$ and the tricritical point via the additional condition that  $\partial^4 G/\partial m^4=0$ at $m=0$. This gives the equation of the zero field critical line of the BC model as 
\begin{equation}
D=\frac{\log 2 (\beta -1)}{\beta }\hspace{3cm}(\mbox{BC model}, H=0)
\label{critBC}
\end{equation}
and
\begin{equation}
\beta_{tcr}=3\,\,\,;\,\,\,D_{tcr}=\frac{\log 4}{3}=0.4621\hspace{1cm}(\mbox{ BC tricritical point})
\label{tcrp}
\end{equation}

Eq.(\ref{critBC}) should be used only for $\beta\le 3$. For higher $\beta$ values while the equation still gives critical values it is for the metastable normal phase which we shall not pursue here. It can also be verified that the zero field coexistence line starting at the tricritical point 
touches $T=0$ at $D=0.5$. For higher values of $D$ the superfluid state no more remains globally stable at any temperature, though it remains locally stable. In other words, for $D>0.5$ the normal state remains the preferred state for all temperatures down to zero. From eq.(\ref{critBC}) it may also be easily checked that for $D=0$, which is just the three state Ising model, the critical temperature is $\beta=3/2$ and for $D\to-\infty$ in which limit the BC model tends to the two state Ising model the critical temperature $\beta\to 1$ as expected. Also plotted in fig. \ref{phase1}(b) is the phase diagram in the $x-T$ plane which shows the critical line, the tricritical point and the also the coexistence region. Here, $x=1-q$ refers to the specific concentration of the $\mbox{He}^3$ phase.

\begin{figure}[t!]
\begin{minipage}[b]{0.3\linewidth}
\centering
\includegraphics[width=2in,height=1.7in]{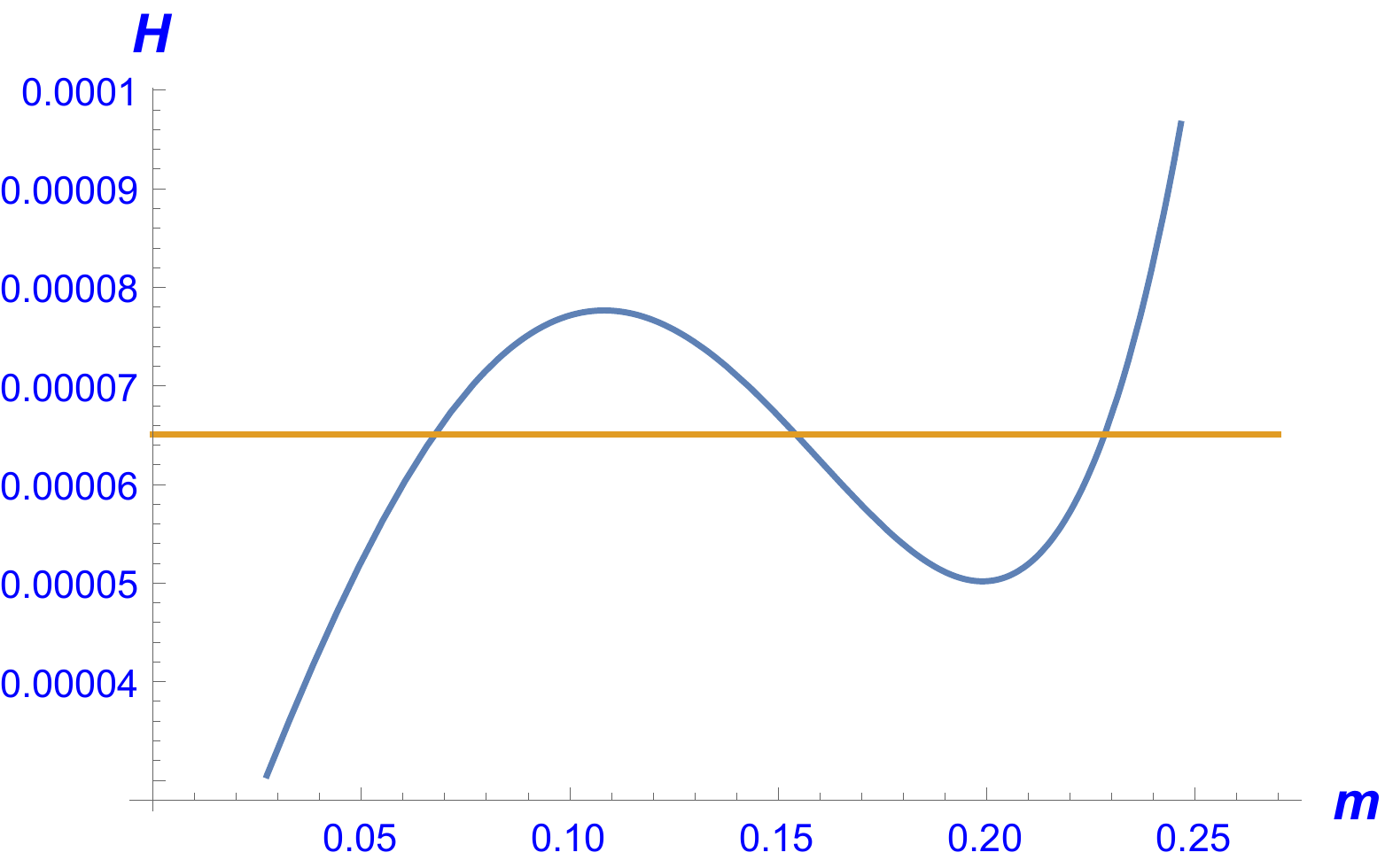}
\end{minipage}
\hspace{0.2cm}
\begin{minipage}[b]{0.3\linewidth}
\centering
\includegraphics[width=2.1in,height=1.8in]{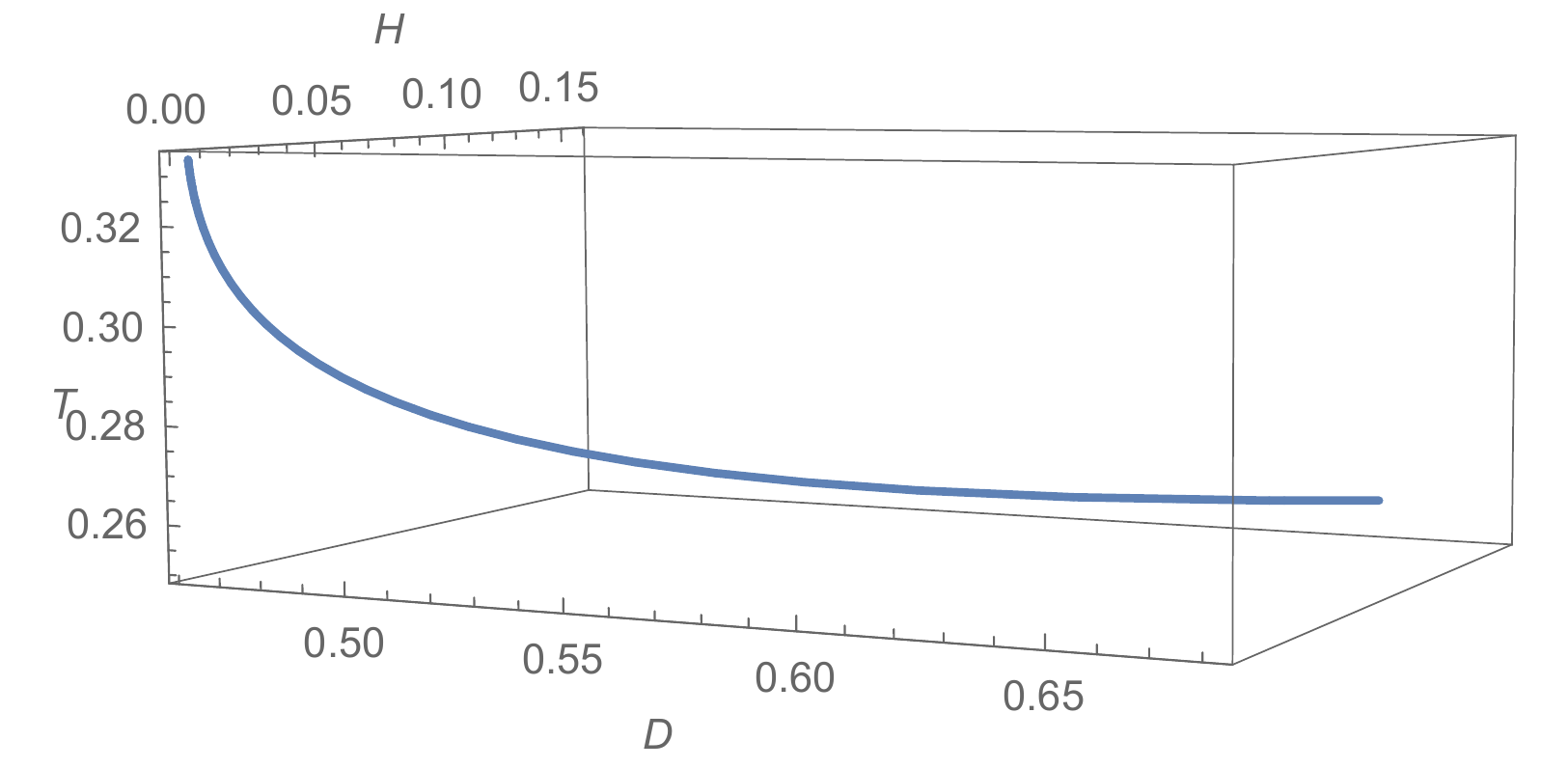}
\end{minipage}
\hspace{0.3cm}
\begin{minipage}[b]{0.3\linewidth}
\centering
\includegraphics[width=1.9in,height=1.8in]{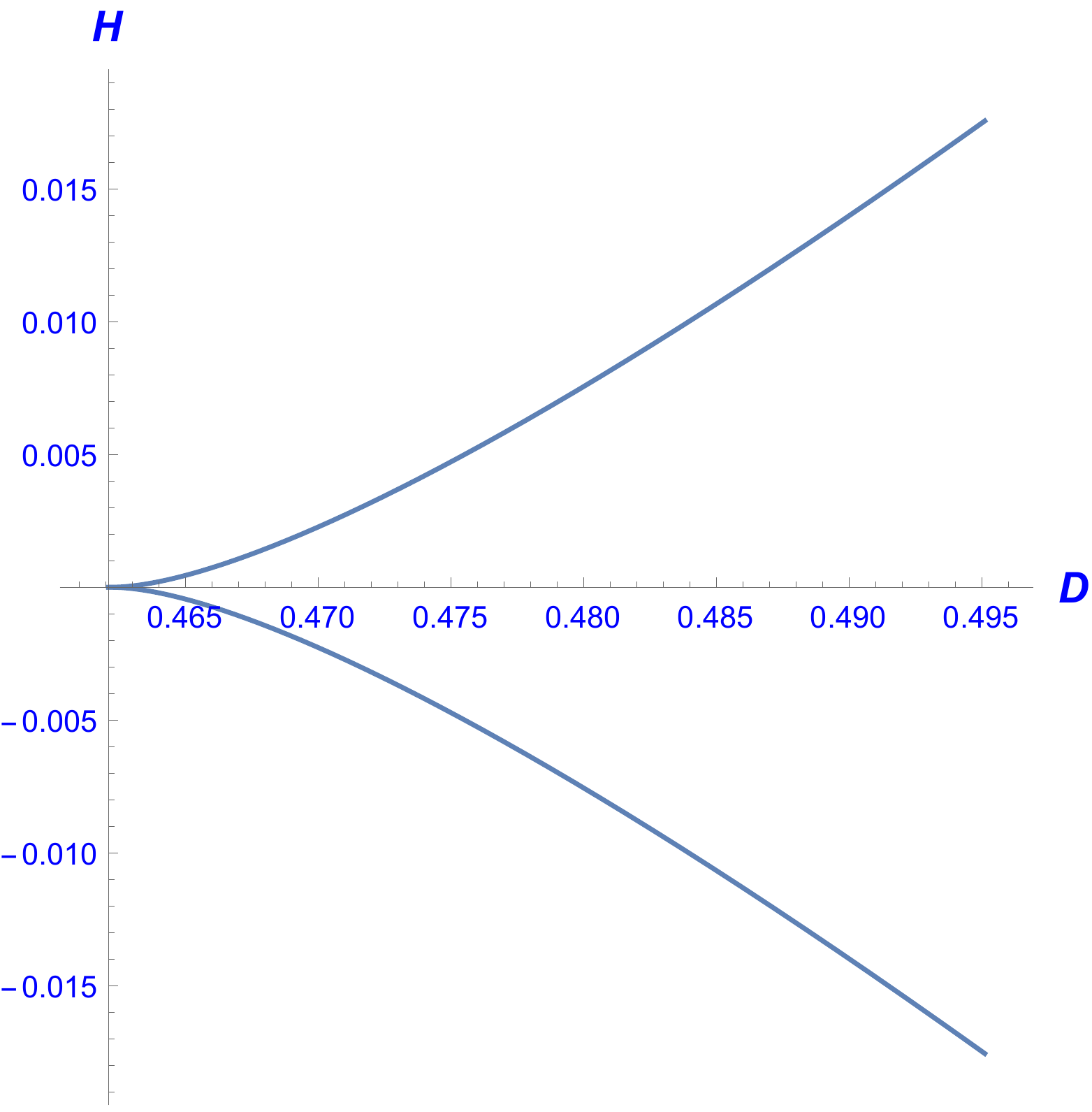}
\end{minipage}
\caption{\small{$(a$). Isotherm in the $H-m$ plane at $D=0.4634,\,\beta=3.08$ shows phase coexistence via Maxwell construction at $H=0.000065$.  $(b)$ The non-zero field critical line bordering from above the coexistence region or the ``wing'' in the $D-H-T$ plane for the Blume-Capel model with $\omega=0$. $(c)$ The projection of the same critical line in the $D-H$ plane. The leftmost point on the $D$ axis is the tricritical point with $D=0.462,H=0$ and $T=1/3$. }}
\label{wingphase}
\end{figure}

Moving further, we now briefly review the phase structure for the non zero field for the BC case. Again, the phase structure is similar for the BEG case with small $\omega$ values. For non zero field it is best to invert eq.(\ref{magnetic moment}) for the magnetic moment and express the magnetic field $H$ in terms of $m$,
\begin{equation}
H=\frac{1}{\beta}\log \left(\frac{m\,e^{\beta  D}+\sqrt{m^2 e^{2 \beta  D}-4 m^2+4}}{2 (1-m)}\right)- m
\label{magfield}
\end{equation}
Interestingly, now this becomes a closed form equilibrium relation. A typical isothermal plot of $H$ vs. $m$ is shown in fig.\ref{wingphase}(a). The curve is analogous to the $P-v$ isotherm in the van der Waals gas and the phase transition point is similarly obtained by drawing an equal area Maxwell construction. On further lowering the temperature the the isotherm will cross the critical point via an inflection given by $\partial^2H/\partial\,m^2=0$, which in turn obtains the critical value of $m$ in terms of $\beta$ and $D$,
\begin{equation}
m_{cr}=\frac{\sqrt{e^{2 \beta  D}-16} \left(e^{2 \beta  D}-4\right)}{\sqrt{192 e^{2 \beta  D}-48 e^{4 \beta  D}+4 e^{6 \beta  D}-256}}
\label{mcritbd}
\end{equation}
Further, on taking the second derivative of $G$ in eq.(\ref{free energy}) with respect to $m$ at $m=m_{cr}$ in eq.(\ref{mcritbd}) above and then plugging in the value of critical $H$ from eq.(\ref{magfield}) we can obtain simple relations between the parameters $D,\,H$ and $\beta$ along the non-zero field critical line, \cite{beg},
\begin{eqnarray}
H_{cr}&=&\frac{1}{\beta}\log \left(\frac{\beta +\sqrt{(\beta -3) \beta }-2}{\sqrt{4-\beta }}\right)-\frac{1}{\beta}\sqrt{(\beta -3) \beta }\nonumber\\
D_{cr}&=&\frac{1}{2\beta} \log \left(\frac{16}{4-\beta }\right)\nonumber\\
m_{cr}&=&\frac{\sqrt{\beta -3}}{\sqrt{\beta }}
\label{wingcriticals}
\end{eqnarray}
It may be easily ascertained from the equation above that along the non zero field critical line (or the `wing' critical line) $\beta_{cr}$ ranges from $3$ to $4$ as $D_{cr}$ ranges from $D_{tcr}$ to infinity and $H_{cr}$ ranges from zero to infinity. In fig. \ref{wingphase}(b) we plot the non zero field critical line in the $D-H-T$ space. The critical line borders the coexistence `wing' under it, one each symmetrically placed in the positive and negative $H$ direction. In fig. \ref{wingphase}(c) the symmetrically placed wings are projected upon the $D-H$ plane.

Admittedly, we have left unexplored much of the parameter space which in fact yields very rich phase structure. For example, at $\omega=3.1$ there are two more tricritical points symmetrically positioned on the critical boundaries of the wings ${\bf B}$ and ${\bf B'}$, \cite{beg, mukamel}. Moreover negative values of the couplings $J$ or $K$, significantly change the phase behaviour in the spin one model, \cite{hoston,wang}. We shall not pursue the cases of anti-ferromagnetic spin coupling or a repulsive biquadratic coupling in this work. While we believe the geometry of these cases will worth investigating, it will also be more subtle due to the presence of staggered spin and quadrupolar orders. We hope to return to these exciting cases in the future.

\subsection{Constructing the geometry of the mean field BEG model}
\label{haha}

We now discuss the thermodynamic geometry of the mean field BEG model by first accounting for the relevant stochastic variables present in the full model as well as in its mean field approximation. To that end we first write the full Hamiltonian eq.(\ref{BEG}) in terms of the stochastic variables
\begin{equation}
\mathcal{ H}_{\,\small{\,beg}}=\mathcal{ F}_1+\mathcal{ F}_2+\mathcal{ F}_3
\end{equation}
where
\begin{eqnarray}
\mathcal{ F}_1 &=& -J\, \sum_{<ij>} \, S_i\, S_{j} - K\, \sum_{<ij>} S_i^{\,2}\, S_{j}^{\,2}\nonumber\\
\mathcal{ F}_2 &=& -H\, \sum_{i} \, S_i\nonumber\\
\mathcal{ F}_3 &=& D\,\sum_{i} S_i^{\,2}
\label{stochastic beg full}
\end{eqnarray}
 
 The mean values of the random variables are related to the extensive thermodynamic quantities,
\begin{equation}
\langle\mathcal{F}_1\rangle=U\,\,;\,\,\langle\mathcal{F}_2\rangle=-H\,M\,\,;\,\,\langle\mathcal{F}_3\rangle=D\,Q\,\,
\end{equation}
where $U$ stands for the internal energy (explained below), $M$ for the total magnetization and $Q$ for the total quadrupole moment. The three random variables are independent (though correlated) implying that in the parameter space there are three independent directions of equilibrium fluctuations. Thus, for example, there exist different values of the total magnetic moment $\sum S_i$ for the same value of total quadrupole moment $\sum S_i^2$ and vice versa.
 The random variable $\mathcal{ F}_1$ contains information about the correlations among spins  and the quadrupole moments. Here $U$, the mean of  $\mathcal{ F}_1$ is the `internal' energy of the BEG system arising out of spin and quadrupole interactions, while the full energy $E=\langle\mathcal{H}_{{\small{beg}}}\rangle$ includes the interaction of the BEG system with the sources of external fields. While one can work with either energies we shall find it convenient to use $U$. 
 
 From the fundamental relation $S(U,M,Q)$ one can in principle obtain the three dimensional thermodynamic metric as the Hessian matrix of the entropy. More directly, one can also obtain the inverse thermodynamic metric by double differentiating the Massieu function  $\psi(\beta,\nu,\mu)$ obtained from the partition function of the full Hamiltonian, eq.(\ref{BEG}), where it is expressed as a function of the entropic intensive variables $\beta$, $\nu=\beta H$, $\mu=\beta D$. The thermodynamic quantities per spin $u,m,q$ are then obtained as the first derivatives of the Massieu function, while the second derivatives constituting the thermodynamic metric give the second moments of fluctuations in $u,m$ and $q$, including their crossed fluctuations.

Except for the one dimensional spin one model, \cite{krinsky}, and a few special two dimensional cases it is not possible to solve for the partition function. On the other hand the mean field partition function is easily solved and, as the Monte Carlo simulations bear out, its phase structure is qualitatively correct, \cite{jain}. Therefore it is worthwhile exploring the geometry of the mean field model. Unlike the full Hamiltonian the mean field Hamiltonian loses all direct information about the spin-spin correlations and retains only the average effect of interactions. Expectedly, this brings down the number of independent stochastic variables from three to two. We rewrite  (scaled) eq. (\ref{meanfieldbegeq}) for the mean field Hamiltonian to delineate the three random variables,
\begin{eqnarray}
  \mathcal{ H}_{\,\small{\,beg}}^{\small{mf}}&=&{-(\frac{ m}{2}\sum_{i} \, S_i+\frac{\omega\,q}{2}\sum_{i} \, S_i^2)}-{H\sum_{i} \, S_i}+{D\sum_{i} \, S_i^2,}\nonumber\\
 &=& \mathcal{F}_1^{mf}+  \mathcal{F}_2+ \mathcal{F}_3
 \label{meanfieldbegeqscaled}
  \end{eqnarray} 
  where the random variable $ \mathcal{F}_1^{mf}$ corresponding to the mean field internal energy is now a linear combination of the random variables $ \mathcal{F}_2$ and $\mathcal{F}_3$ defined previously in eq.(\ref{stochastic beg full}). With its equilibrium fluctuations governed by the fluctuations and correlations in  $\mathcal{F}_2$ and $\mathcal{F}_3$ the internal energy random variable $\mathcal{F}_1^{mf}$ carries no additional information about the underlying system unlike its full BEG counterpart in eq.(\ref{stochastic beg full})\footnote{The geometry of the isotropic BEG model, with $H=D=0$, was worked out in \cite{erdem} using a metric different from ours.  }. 
We emphasize here that while on the one hand the parameter space of the $BEG$ model is three dimensional with independent $T, H$ and $D$ directions on the other hand at each point in the thermodynamic manifold of the {\it{mean field}} $BEG$, the tangent space of fluctuating directions is only two dimensional. We believe that it is generally true that the number of independent stochastic variables in any mean field approximation of any model is at least one less. 

 This is similar to the case of the two state Ising model, to which we digress for a bit, where the mean field Hamiltonian has only one independent random variable, \cite{mrugala,sarkar1}, as opposed to the full Hamiltonian which has two random variables. This can be seen by recalling the equation of the Ising model Hamiltonian,
\begin{equation}
\mathcal{H}_{ising}=-J\sum_{<ij>}\,S_i\,S_j-H\sum_i\,S_i
\label{ising}
\end{equation} 
and that of its mean field approximation after rescaling by $Jz$ as
\begin{equation}
\mathcal{H}_{ising}^{mf}=-(H+\frac{m}{2})\sum_i\,S_i
\label{meanfieldising}
\end{equation}
Thus, while the parameter space is two dimensional with $T$ and $H$ as independent directions, and so is the thermodynamic manifold of the full Hamiltonian in eq.(\ref{ising}) with two random variables related to the internal energy and magnetization, the thermodynamic manifold of the mean field approxiamtion in eq.(\ref{meanfieldising}) degenerates to a one dimensional geometry with the magnetization as the only independent stochastic variable. This can also be seen by examining the expression for the mean field entropy of the 2-state Ising model.
\begin{eqnarray}
S &=& \log (2 \cosh (m (\beta +H)))-\beta  m (H+m)\nonumber\\
&=& \log \frac{2}{\sqrt{1-m^2}}-m \tanh ^{-1}m
\label{entropy Ising2}
\end{eqnarray}
where in the second equality of eq.(\ref{entropy Ising2}) above we have used the expression for $H$ obtained by invering the self consistent expression for magnetization of the 2-state Ising model. It is clear from the second equality that the entropy depends only one one extensive variable thus rendering the geometry degenerate. 

Similarly, the mean field geometry of the spin one Ising model, with $S=-1,0,1$ is also degenerate as can be quickly ascertained by setting $\omega$ and $D$ to zero in the mean field Hamiltonian, eq.(\ref{meanfieldbegeqscaled}). The resulting mean field Hamiltonian for spin one Ising model once again depends on only one stochastic variable $\sum S_i$, thus rendering the geometry of fluctuations one dimensional. Analogous to eq.(\ref{entropy Ising2}) the entropy for the mean field spin one Ising model can be as a exppressed as a function only of magnetization,
\begin{equation}
S=\log \left(2 \cosh \log \frac{\sqrt{4-3 m^2}+m}{2 (1-m)}+1\right)-m\log\frac{\sqrt{4-3 m^2}+m}{2 (1-m)}
\label{entrop spin one ising}
\end{equation}

 We return now to the mean field BEG model. With two independent random variables the mean field BEG model in eq.(\ref{meanfieldbegeq}) or (\ref{meanfieldbegeqscaled}) permits a non-trivial state space Riemannian geometry. Working with the one dimensional spin-one model in paper I we had explicitly worked out its non-degenerate three dimensional Riemannian state space geometry and had also obtained two sectional curvatures $R_m$ and $R_q$ which were found to closely follow the calculated correlation lengths in order parameters $m$ and $q$ respectively. The sectional curvature $R_m$ was defined on the constant $D$ plane in the $T$-$D$-$H$ parameter space while the scalar curvature $R_q$ was defined on the constant $H$ plane in the parameter space. In this context we stress that the description of an geometry on a hypersurface in the parameter space is physically equivalent to restricting all the equilibrium fluctuations on that hypersurface, with its intrinsic metric induced from the ambient metric \footnote{ See \cite{sahay restricted} for a detailed exposition where the concept was first introduced within the setting of black hole thermodynamics. In \cite{reiksh} the concept was discussed again in the context of the one dimensional spin one model.}. It was also argued and subsequently demonstrated in \cite{reiksh} that the fluctuations in the quadrupole moment are somewhat suppressed in the $D$ plane while those in the magnetization are similarly diminished in the $H$ plane so that these hypersurfaces are good starting points for exploring separately the correlations in the magnetization and the quadrupole moment respectively. On the other hand we also discussed that it was not a good idea to constrain spontaneous fluctuations on surfaces of constant $m$ or $q$ since it would lead to contrived fluctuations.

As stated earlier, at any given point in the three dimensional parameter space of the mean field spin one model there are only two independent directions of equilibrium fluctuations. This means that if we take an arbitrary two dimensional equilibrium hypersurface in the three dimensional state space and in principle constrain the system to undergo spontaneous fluctuations only on the given hypersurface then the thermodynamic metric defined on it will set up a two dimensional Riemannian manifold. The thermodynamic metric on the hypersurface is straightforwardly obtained by taking the derivatives of the entropy (or the free energy) only along directions tangent to the hypersurface. In a sense, therefore, each hypersurface carries its own two dimensional Riemannian geometry, since there is no ambient three dimensional metric to induce projection metrics on these hypersuraces. Of course such a construction would break down in places where one of the directions of independent fluctuations happens to be orthogonal to the hypersurface. In such cases there would be families of curves on the hypersurface along which the entropy remains constant, so that the intrinsic geometry on the surface degenerates to a one dimensional geometry. We shall encounter such cases in the sequel. 

Following our discussion on the geometry of the one dimensional model here too we shall pursue the intrinsic thermodynamic geometries on the $H$ surface and the $D$ surface. Similar to the one dimensional case, we find it easier to work with the Massieu functions which we could express from eq.(\ref{massieu}) either as $\psi_D(\beta,\nu)$ with $D$ constant or as $\psi_H(\beta,\mu)$ with $H$ constant. The corresponding entropy representations are as $S(U_1,M)$ or as $S(U_2,Q)$ where, it can be checked, the enthalpy like quantities are 
\begin{eqnarray}
 U_1&=&U+Q\,D\nonumber\\
 U_2&=&U-H\,M
 \label{enthalpy}
 \end{eqnarray}
wherein $D$ is to be held constant in the former and $H$ in the latter\footnote{Note that while we are using the extensive quantities $S,U,M$ and $Q$ in our discussion in this subsection, elsewhere in the paper we refer only to their specific values obtained by dividing by the total number of spins $N$.}. In both cases the Massieu function is the same but the different arguments are emphasised here since the Riemannian metrics will be built out of the double derivatives in the arguments. It is helpful to write the thermodynamic laws relevant to the first and the second geometries,
\begin{eqnarray}
T\,dS&=&dU_1-H\,dM\nonumber\\
d\psi_{\small{D}}&=&-U_1\,d\beta-M\,d\nu\hspace{1in}(D\, \mbox{constant})
\label{first law 1}
\end{eqnarray}

\begin{eqnarray}
T\,dS&=&dU_2+D\,dQ\nonumber\\
d\psi_H&=&-U_2\,d\beta+D\,d\nu\hspace{1in}(H\, \mbox{constant})
\label{first law 2}
\end{eqnarray}
The metric for each geometry can be obtained via the Hessian of either the entropy or the Massieu function (the inverse metric). It can be expected that in the first case the Riemannian geometry would be more sensitive to the fluctuations in the magnetic moment as compared to the quadrupole moment while in the second case the opposite would be true. As mentioned earlier this is strongly suggested by the geometry of the one dimensional spin one model \cite{reiksh}, and we shall sometimes refer to the first geometry which restricts fluctuations to the constant $D$ plane in the $T$-$H$-$D$ parameter space as the $m$-geometry and the second one which limits fluctuations to the constant $H$ plane as the $q$-geometry. We accordingly designate the metric and the state space scalar curvatures from the $m$-geometry as the $m$-metric and $R_m$ and similarly from the $q$-geometry as the $q$-metric and $R_q$. 

It is immediately obvious from eq.(\ref{meanfieldbegeqscaled}) that the $q$-geometry becomes degenerate in the normal phase with $m=H=0$. This is because with the $q$-geometry fluctuations restricted to the $H=0$ plane all the coefficients of the stochastic variable $\sum_iS_i$, including the variable itself, vanish leaving only a single random variable $\sum_iS_i^2$ free to fluctuate in the plane. On the other hand the $m$-geometry, which is defined on the $D$ plane, continues to remain non degenerate in the normal phase since any fluctuation orthogonal to the $H$ plane amounts to a non zero $m$ as well as $H$, so that both the stochastic variables need to be retained. A somewhat limiting consequence of this is that in the normal phase there is only the curvature $R_m$ to guide us. However, everywhere else in the parameter space (except the line $D=0,\omega=0$ as already stated) both the curvatures inform the underlying correlations in the two order parameters. We mention here that possibly there could be a way to resurrect the $q$ geometry in the normal phase by adding a lattice dependent energy term to the Hamiltonian (\ref{meanfieldbegeqscaled}). We shall not be pursuing this line here but refer to \cite{mrugala} and \cite{sarkar1} where such an approach was used in the context of the mean field Ising model.

\section{Geometry of criticality}
\label{hu}

 We now probe the scaling behaviour of the state space scalar curvature in light of the Ruppeiner equation which relates the curvature to the inverse of the critical free energy. 
  With its three dimensional parameter space the singular free energy of the BEG model could depend on three independent scaling fields obtained from a linear combination of deviations of the magnetic field $H$, the anisotropy field $D$ and the temperature $T$ from their respective critical values. For the full BEG model eq.( \ref{BEG}) this is indeed the case as is worked out thoroughly in \cite{lawrie}, especially near the tricritical point.  
We are unable to ascertain if this will hold out in the mean field context, given that the number of random variables here is reduced to two. In any case since the focus of our investigation is the geometry in either the $H$-plane or the $D$-plane we shall restrict the scaling analysis to these planes only, so that the critical free energy here shall have only two scaling fields.
 
  Following \cite{rupprev} we first briefly summarize the context for obtaining the asymptotic form of scalar curvature. The scaling form of the singular free energy density can be expressed in terms of the scaling fields $t=(\beta_c-\beta)/\beta_c$ and the ordering field $h$ as,
 \begin{equation}
\omega(t,h)= n_1 \,t^{2-\alpha}\,Y(n_2\, h\, t^{-\widetilde {\beta} \delta})
\label{scaling free}
\end{equation}
where the universal critical exponents $\alpha,\widetilde{\beta}$ and $\delta$ have their usual meanings as critical exponents while the constants $n_1$ and $n_2$ are non universal and system dependent\footnote{To avoid confusion we shall refer to the ``critical exponent $\beta$'' by the symbol $\widetilde\beta$ while reserving the symbol $\beta$ for the inverse temperature.}. $Y(z)$ is the spin scaling function which depends on $h$ and $t$ in a single argument combination $z= h t^{-\widetilde{\beta}\delta}$. Using the scaling form of free energy from eq.(\ref{scaling free}) in the evaluation of $R$ and putting it back into the Ruppeiner equation in eq.(\ref{Ruppeiner eq}), one obtains a third order differential equation for the function $Y(z)$ which in turn leads to the scaling form of $R$, \cite{rupprev},
\begin{equation}
R=\frac{\widetilde{{\beta}}(\widetilde\beta\delta-1)(\delta-1)k_B\,T_c}{(2-\alpha)(1-\alpha)Y(0)}\, t^{\alpha-2}
\label{Rscaling}
\end{equation}
On the other hand the constant field specific heat $C_h$ can be evaluated directly from eq.(\ref{scaling free}) and its leading singular part in zero field is
\begin{equation}
C_h= -\frac{(2-\alpha)(1-\alpha) Y(0) t^{-\alpha}}{T_c}
\label{specific heat scaling}
\end{equation}
The product of the scaling forms of $R$ and $C_h$ from equations (\ref{Rscaling}) and (\ref{specific heat scaling}) given as
\begin{equation}
RC_h t^2= -\widetilde{\beta}(\delta-1)(\widetilde{\beta} \delta-1) k_B
\label{RChtsquare}
\end{equation}
is consistent with the conjectured correspondence of $R$ with the correlation volume $\xi^d$ (and hence with the inverse free energy via hyperscaling). This follows from a well known prediction of two scale factor universality according to which the product similar to the one in eq.(\ref{RChtsquare}) above, with $R$ replaced by the correlation volume $\xi^d$ is shown to be equal to a constant which depends only on the universality class, (\cite{rupprev},\cite{stauffer}).

In the following we shall present our main results for the geometry near the zero field critical line, tricritical point, near the wing critical line. But first we pause briefly to make some general comments in favour of the geometric correlation length obtained via the state space scalar curvature.

\subsubsection{Geometric vs. the Ornstein-Zernike correlation length: a divertissement}
\label{hai}

It shall be the main task of this paper to make a case that the curvatures $R_m$ and $R_q$ faithfully represent, to a large extent, the correlation lengths in the order parameters $m$ and $q$ respectively. We repeat that unlike the one dimensional case where the correlation length is exactly calculable via the transfer matrix and hence easily comparable to its geometric counterpart, there is no direct way to determine correlation length in a mean field set up since we have already averaged out the effect of spin-spin interactions here. The standard way of getting around
this limitation is to add a square gradient term as a first correction to the mean field Hamiltonian, assuming a slowly varying order parameter, \cite{pathria}. From the resulting Landau-Ginzburg Hamiltonian we can obtain a correlation length which scales as $\xi\sim t^{-1/2}$ at $h=0$ irrespective of the dimension or the coordination number. Thus the exponent $\nu=1/2$ and the upper critical dimension for ordinary critical points in $d=4$.

 We recall that the geometric method of estimating the scalar curvature is distinctly different. The curvature length scale in the state space manifold is the one {\it beyond} which the local flatness does not hold, and this translates in the thermodynamic system to the physical length scale {\it below} which the local correlation effects are so strong that the mean field approximation breaks down, \cite{rupprev}. Therefore, as alluded to earlier, scalar curvature is an {\it in-built measure} of the length scale at which the classical fluctuation theory breaks down and uses only the information already available from thermodynamics. In this work for the BEG model, as for the van der Waals case earlier, \cite{rupprev,widom}, the universality class remains the Ising uiversality class with well known critical exponents obtained theoretically. As we shall also see in the sequel the scalar curvature $R_m$ is always found to scale as $t^{-2}$ near the critical point for models which belong to the Ising class. Given that the state space scalar curvature scales as the correlation volume $\xi^d$ we can obtain the exponent $\nu$ from geometry in a straightforward manner. In the following table we compare the Ising, the geometric and the mean field  critical exponent $\nu$ for different dimensions
 
 \begin{table}[h!]
\centering

 \begin{tabular}{c c c c } 
 \hline
dimension & true $\nu$ & geometric $\nu$ & mean-field $\nu$ \\ 
 \hline
2 & 1 & 1 & 0.5 \\ 
 
 3 & 0.630 & 0.667 & 0.5\\
 
 4 & 0.5 & 0.5& 0.5\\
 \hline
 \label{table}

\end{tabular}
\label{table correlation}
\caption{\small{Table comparing  in different dimensions the Ising critical exponent $\nu$ obtained via the standard renormalization group calculations (labeled `true'), the ones obtained via thermodynamic geometry (labeled `geometric') and the mean field value.}}
\end{table}
While all this should certainly not be taken as a suggestion of any superiority of the geometric correlation length over the  square gradient one (the latter has a very well founded basis and well established usefulness), it certainly does help substantiate the Ruppeiner conjecture relating the state space scalar curvature to the correlation length, especially near criticality. Additionally, we hope it could also inform the RG calculations, \cite{sarkar reno} and, coupled with the fact that the geometric curvature is straightforward to calculate, it could also help speed up such calculations. Significantly, as was pointed out in \cite{sahay1} in the context of simple fluids, the association of $R$ with the correlation length could go far beyond the neighbourhood of the critical region. In fact we shall have ample opportunity in the sequel to confirm that, consistent with Widom's arguments equating the correlation lengths across the interface, the scalar curvatures in the coexisting phases reasonably agree with each other in places far enough from criticality.

In the following sections we investigate, in turn, the thermodynamic geometry near the critical points and near the coexistence points, both in the zero field and in the non-zero magnetic field.

\subsection{Geometry near the zero field critical line $\lambda$}
\label{hai1}

\subsubsection{the $m$-geometry}
We first discuss the Blume-Capel case ($\omega=0$), for which we are able to obtain relatively tractable algebraic expressions. In order to obtain a scaling expression of the scalar curvature $R_m$ we first find its value in the normal phase with $m=0$,
  \begin{equation}
  R_m^{-1}=-\frac{2 D^2 \left(-2 \beta +e^{\beta  D}+2\right)^2}{e^{\beta  (-D)} \left(e^{\beta  D}+2\right) \left(D e^{\beta  D}-2\right) \left(D \left(-2 \beta +(\beta +1) e^{\beta  D}+2\right)-e^{\beta  D}-2\right)}
  \label{Rmnormal}
  \end{equation}
  We then expand $R_m^{-1}$ in powers of the small parameter $t=(\beta_c-\beta)/\beta_c$ in the neighbourhood of the critical line given by eq.(\ref{critBC}). The dominant term in the expansion is 
\begin{equation}
R_m^{-1}=-2(\beta_c-1) D_c^2\,t^2+O\left(\text{$t^3 $}\right)
\label{Rmscaling}
\end{equation}

On the other hand, the zero field specific heat $C_h$ in the normal phase is obtained as
\begin{equation}
C_h=\frac{2 \beta ^2 D^2 e^{\beta  D}}{\left(e^{\beta  D}+2\right)^2}
\label{chbc}
\end{equation}
and near the critical point it can be expanded as
\begin{equation}
C_h=\,(\beta_c-1)\,D_c^2 +O\left(\text{$t$}\right).
\label{Ch}
\end{equation}
Hence, from eq.(\ref{Rmscaling}) and (\ref{Ch}), the product
\begin{equation}
R_m\,C_h\,t^2=-\frac{1}{2}
\label{rchtsq)}
\end{equation}
 is what one would get by putting in mean field values of exponents in eq.(\ref{RChtsquare}), namely $(\alpha,\widetilde{\beta},\gamma,\delta)=(0,1/2,1,3)$. This establishes the appropriateness of the scalar curvature $R_m$ in encoding the zero field critical behaviour in the BEG model and further strengthens the association of $R_m$ with the correlation volume of fluctuations in $m$, at least in the critical region. To complete the picture we also present the scaling behaviour of $R_m$ with the magnetic field $H$ in the $D$ plane, along the $t=0$ line. It works out to,
 \begin{equation}
 R_m\sim h^{-4/3}\hspace{0.5in}(t=0, \mbox{ crtitical point} )
 \label{Rm hscaling}
 \end{equation}

We turn now to the case of non zero coupling between the quadrupole moments, namely the full BEG case. The algebraic expressions involved are too lengthy so we resort to numerical investigations for fixed values of the ratio $K/J=\omega$ and the neighbourhood of a fixed critical point. With a non zero $\omega$ the zero field quadrupole moment in the normal phase is now obtainable via an implicit relation,
\begin{equation}
q=\frac{2 e^{\beta  q \omega -\beta  D}}{2 e^{\beta  q \omega -\beta  D}+1}
\label{quadzbeg}
\end{equation}
Near the critical point in the normal phase the implicit equation above can be used to express a small deviation of the quadrupole moment from its critical value in terms of the deviation of the temperature from its critical value. It can be checked that the dependence on temperature is linear. Finally, a series expansion in the normal phase of $R_m$ and $C_h$ in the vicinity of, for example, the critical point at $\beta_c=2.239, D_c= 0.45$ with $\omega=0.1$ is given as
\begin{equation}
R_m^{-1}=-0.3451\,t^2 + O(t)^3
\end{equation}
and
\begin{equation}
C_h= 0.2155 - 0.4848\, t + O(t)^2
\end{equation}

We note that while the scaling of $R_m$ is correct the product of amplitudes $R_m C_h t^2$ equals $0.62$ and not the universal value of $1/2$ as expected with mean field exponents and as verified in the case of the Blume Capel limit in the preceding. Nevertheless the product is always found to be greater than half and less than unity . For example, for the critical point at $\beta_c=2.3932,D_c=0.47$ with $\omega=0.1$ the product equals $0.756$ and for $\beta_c=2.0126,D_c=0.45$ with $\omega=0.2$ the product equals $0.537$. Thus it is seen that the product, while still small, seems to depend on the ratio $\omega$ and also on the location of the critical point. We defer further analysis to a future investigation.
\begin{figure}[t!]
\begin{minipage}[b]{0.3\linewidth}
\centering
\includegraphics[width=2in,height=1.5in]{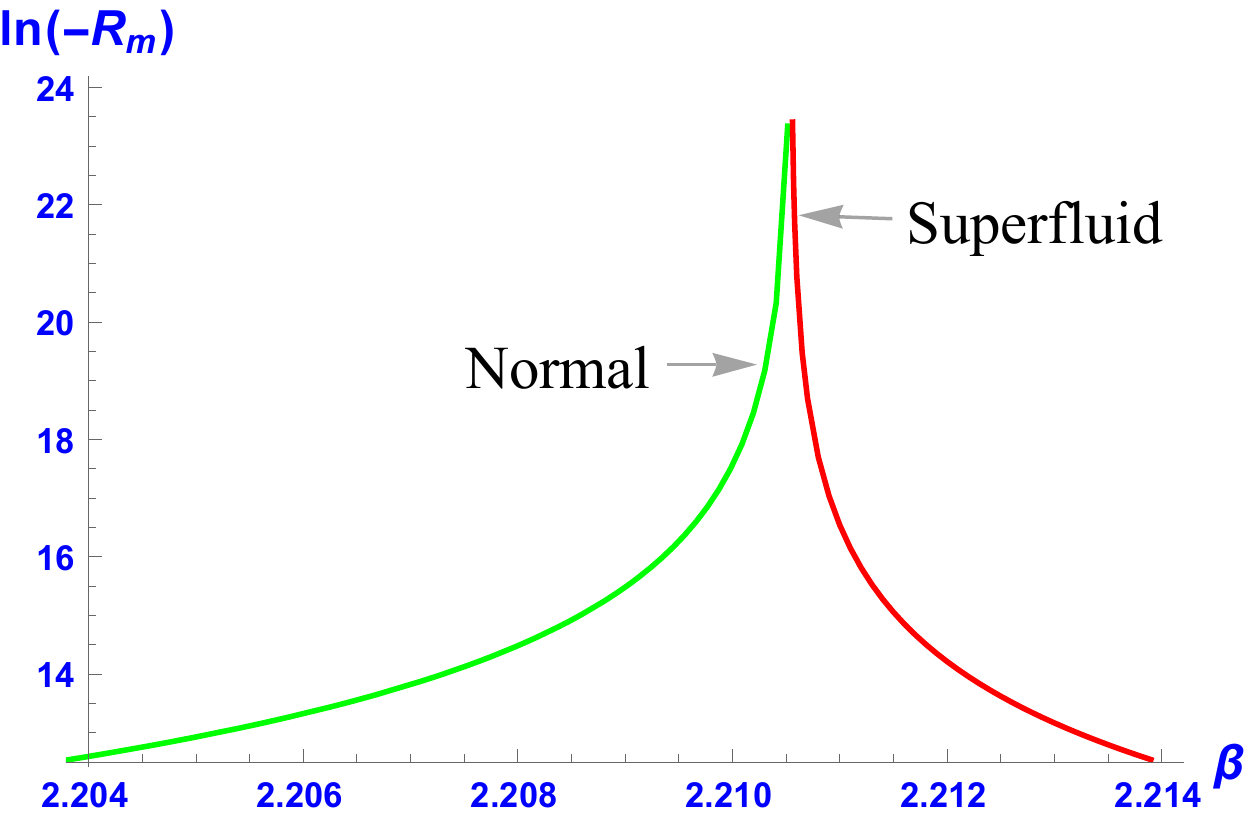}
\end{minipage}
\hspace{0.5cm}
\begin{minipage}[b]{0.3\linewidth}
\centering
\includegraphics[width=2in,height=1.5in]{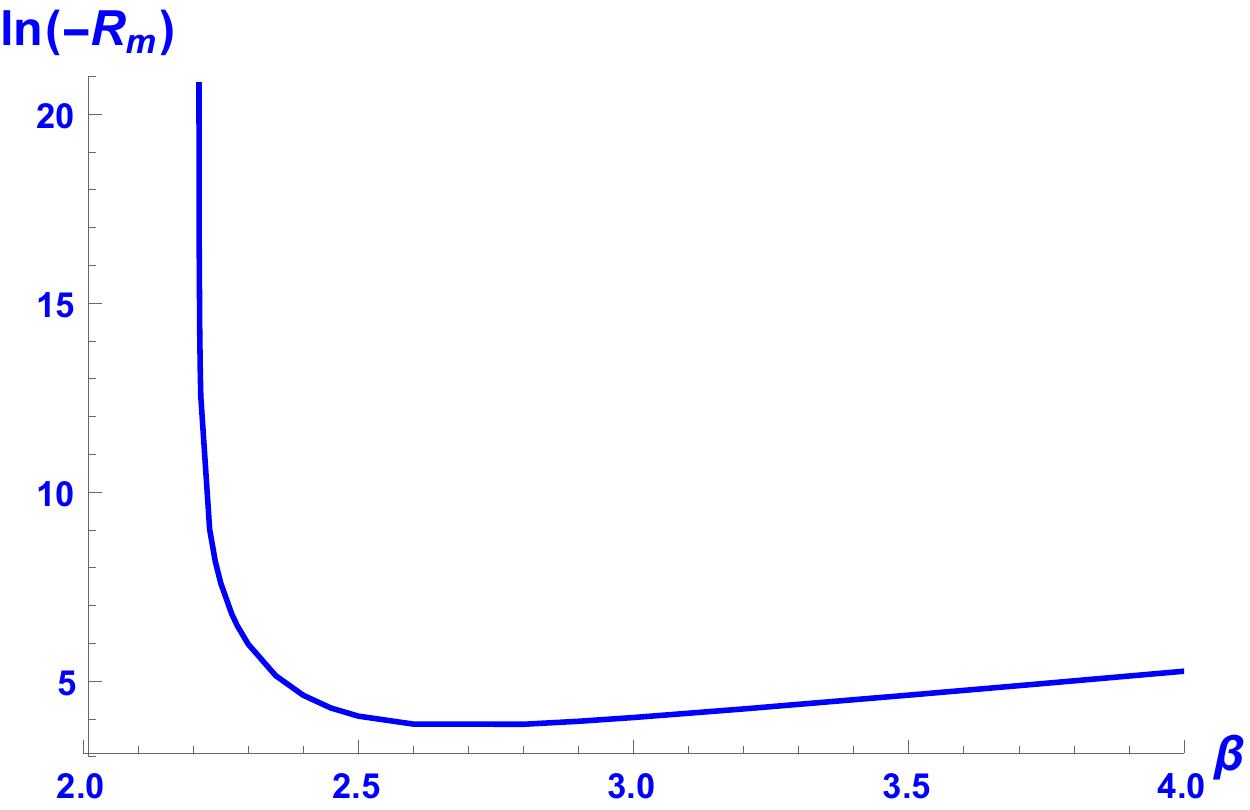}
\end{minipage}
\hspace{0.5cm}
\begin{minipage}[b]{0.3\linewidth}
\centering
\includegraphics[width=2in,height=1.5in]{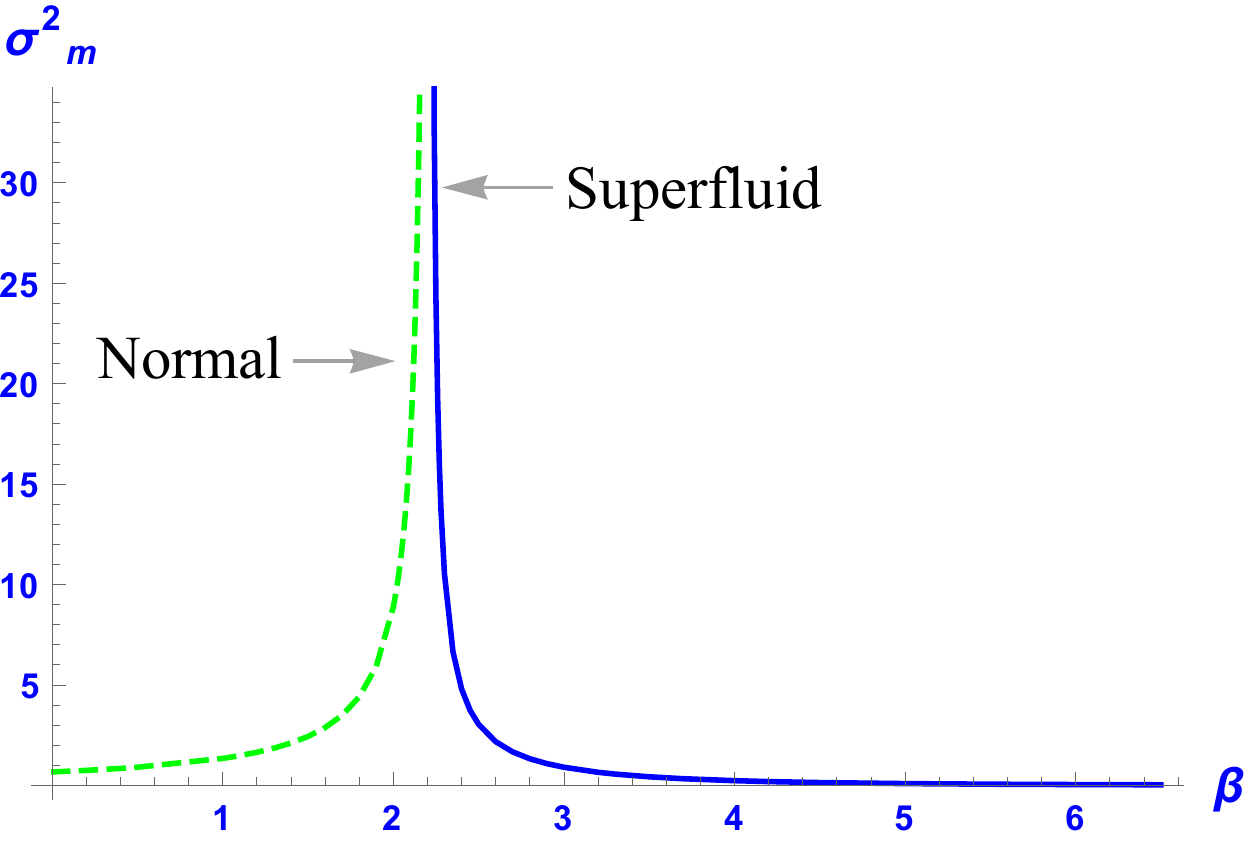}
\end{minipage}
\caption{\small{$(a)$ Plot of $\log(-R_m)$ vs. $\beta$ for the normal and superfluid (or ferromagnetic) phases, in the Blume-Capel limit $\omega=0$. $(b)$  $\log(-R_m)$ vs. $\beta$ for the superfluid phase with $\beta$ ranging from the critical value to higher values (lower temperatures). $(c)$ Plot of the magnetization fluctuation $\sigma _m^2$ vs. $\beta$. In all the sub-figures, the value of $D=0.4$ and the critical point is at $\beta=2.211$.}}
\label{zfcr1}
\end{figure}

We now briefly digress to note some general features of this geometry for the zero field case. In fig.\ref{zfcr1}.(a) we obtain a semi-log plot of scalar curvature $R_m$ vs. $\beta$ across the critical line at $D=0.4$. $R_m$ is seen to diverge to negative infinity from both the normal and superfluid phases. Further extending the semi log plot of $R_m$ into the superfluid phase  we see in fig\ref{zfcr1}.(b) that after dropping to low values sufficiently far away from the critical line the curvature $R_m$ begins a slower divergence to negative infinity on approaching the zero temperature. Finally, in fig(\ref{zfcr1}).(c) the magnetization fluctuation is seen to diverge at the critical point from either side and, it can be established, decays to $2/3$ as $\beta\to0$ and to zero as $\beta\to\infty$ in the superfluid phase, irrespective of $D$.  The second moments in $m$ and $q$, termed  $\sigma^2_m$ and $\sigma^2_q$ here, are obtained via a double partial differentiation of the Massieu function $\psi(\beta,\mu,\nu)$ in eq.(\ref{massieu}) in terms of $\nu$ and $\mu$ respectively. Admittedly, we do not understand the reason for the divergence in $R_m$ near zero temperature since the fluctuations in magnetization decay to zero in the limit of zero temperature thereby suggesting that $R_m$ too decay. We note, however, that such a trend is observed in many black hole thermodynamic systems which have mean field like equations of state and whose state space scalar curvature diverges at extremality (where the horizon temperature goes to zero). Nonetheless, we do see a very good agreement in the trends of $R_m$ and $\sigma^2_m$ around and reasonably further from the critical point.

\begin{figure}[]
\begin{minipage}[b]{0.3\linewidth}
\centering
\includegraphics[width=2in,height=1.5in]{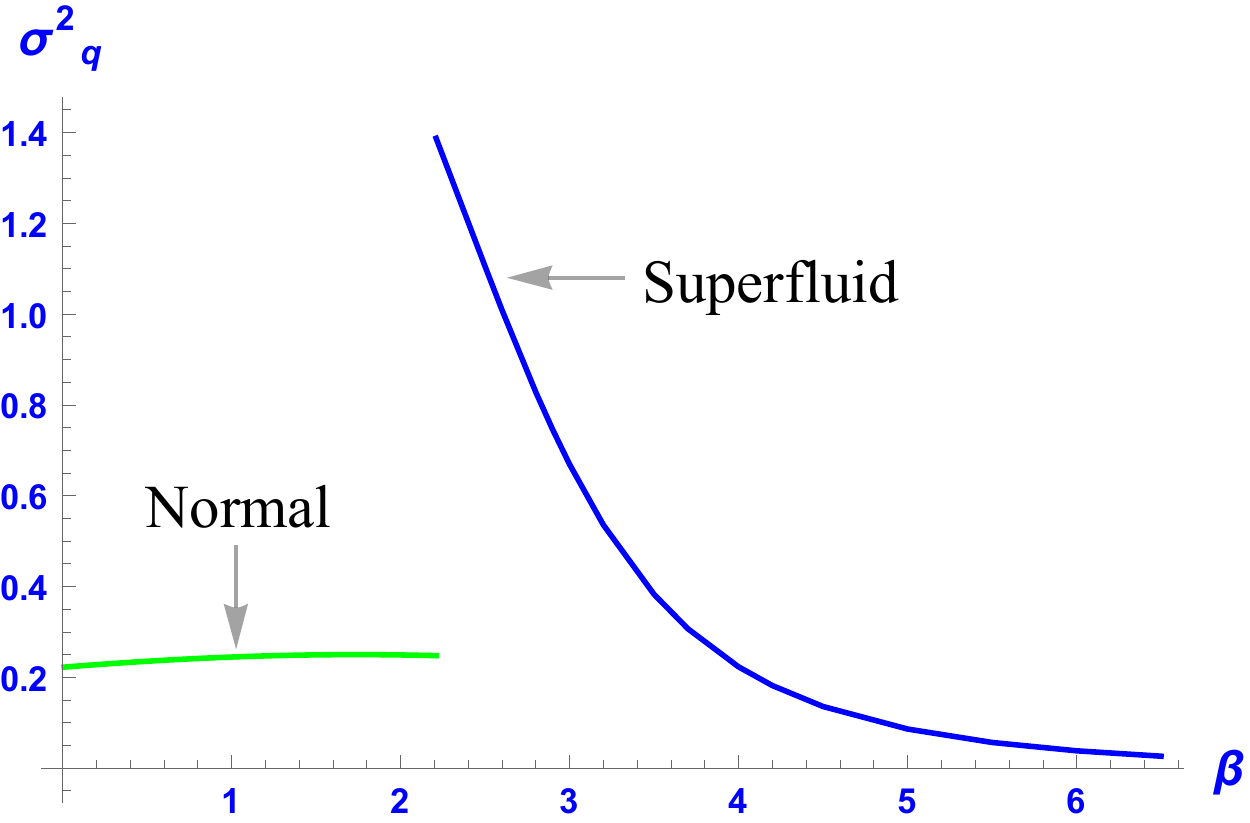}
\end{minipage}
\hspace{0.2cm}
\begin{minipage}[b]{0.3\linewidth}
\centering
\includegraphics[width=2in,height=1.5in]{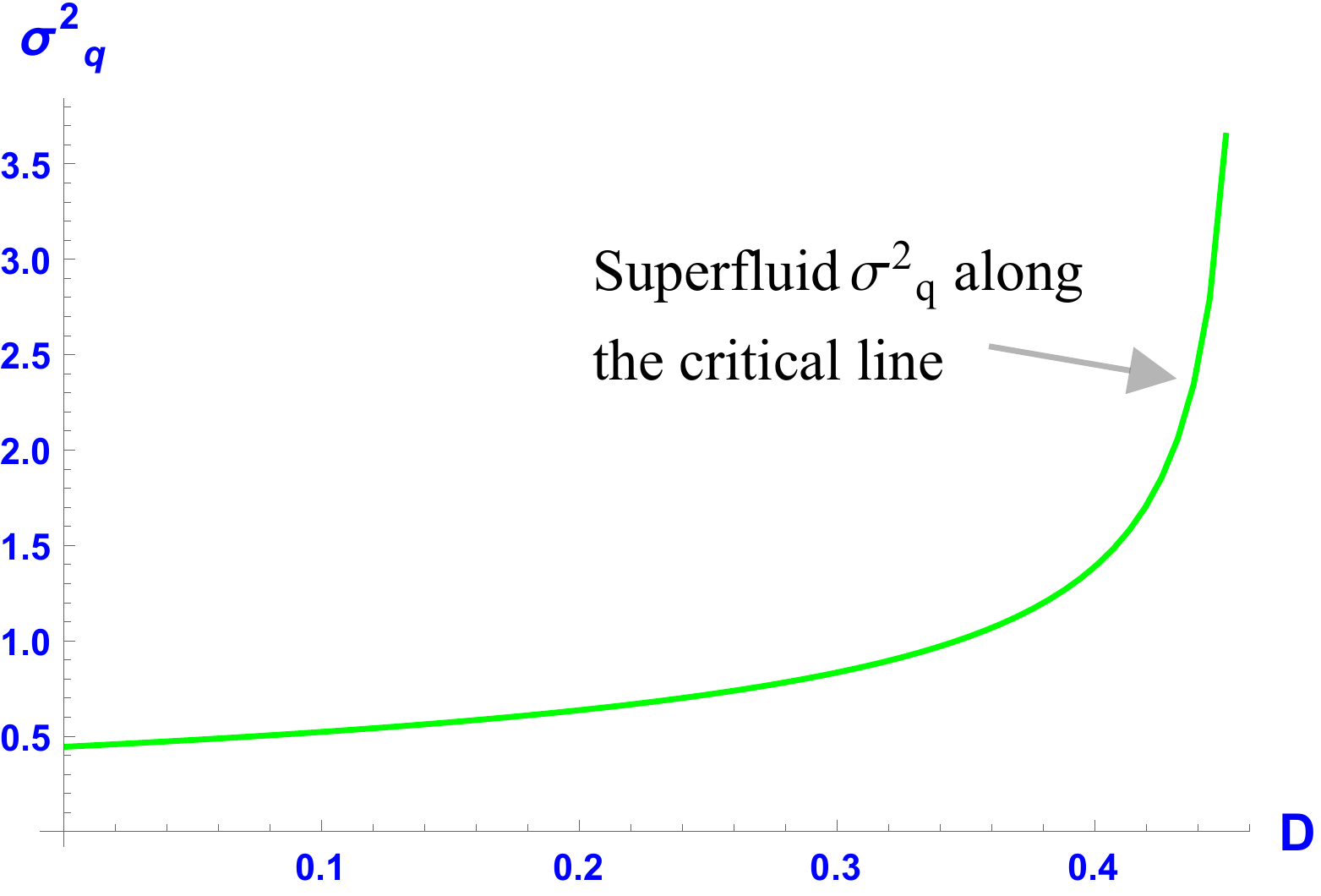}
\end{minipage}
\hspace{0.2cm}
\begin{minipage}[b]{0.3\linewidth}
\centering
\includegraphics[width=2in,height=1.5in]{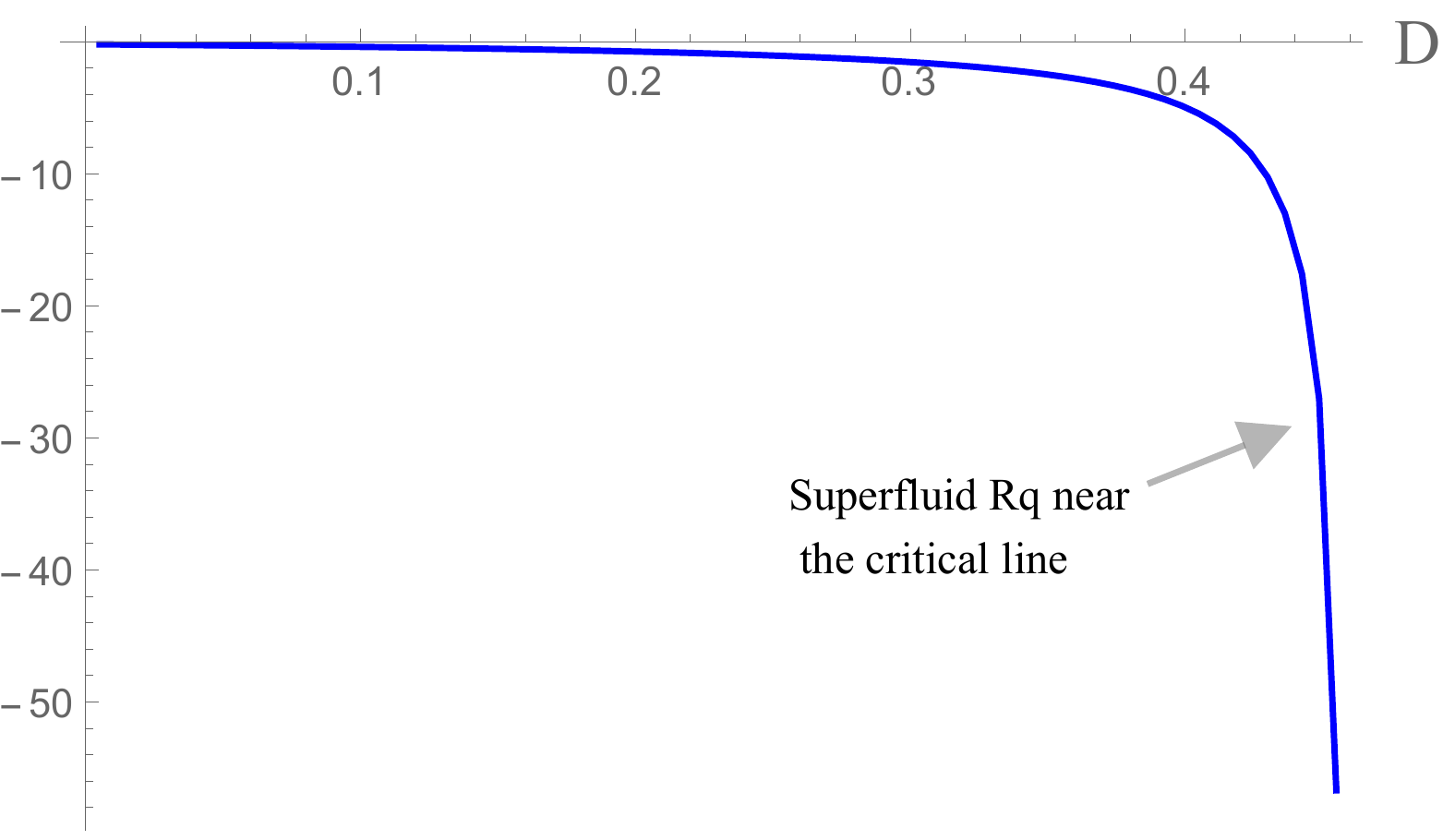}
\end{minipage}
\caption{\small{  $(a)$ Plot of quadrupole fluctuation $\sigma_q^2$ vs. $\beta$ for $D=0.4$ in both the normal and the superfluid phases. Notice the discontinuity at the critical point $\beta=2.211$. Plots {\it along} the zero field critical line of $(b)$ superfluid quadrupole fluctuation $\sigma_q^2$ vs. $D$ and $(c)$ of superfluid scalar curvature $R_q$ vs. $D$. Refer to fig.\ref{phase1}(a) for the zero field critical line. For plots $(b)$ and $(c)$ all points are at the reduced temperature inverse $t=10^{-5}$. In all subfigures $\omega=0$.}}
\label{zfcr2}
\end{figure}

\subsubsection{the $q$-geometry}

As mentioned earlier the $q$ geometry becomes degenerate in the zero field normal phase, as is easily seen in the expression for the normal phase entropy in the Blume-Capel limit,
\begin{equation}
S=q \log \left(\frac{2}{q}-2\right)+\log \left(\frac{1}{1-q}\right)\hspace{1in}(\mbox{Blume-Capel normal phase})
\label{entropy normal}
\end{equation}
While it is not surprising that the normal phase Blume-Capel entropy depends only on $q$ and not on $m$ or $H$, which are both zero there, for the general case of non zero $\omega$ the entropy would also depend on $\beta$. However, the determinant of the $q$-metric can always be shown to be identically zero. It would  be worthwile exploring the nature of the fluctuations $\sigma^2_q$ in both the normal and the superfluid phases. In fig.\ref{zfcr2}(a) we plot $\sigma^2_q$ vs. $\beta$ in the Blume-Capel limit for $D=0.4$ (same as fig(\ref{zfcr1})). Clearly, the quadrupole fluctuations in the normal phase are small but also quite flat in that the $\sigma^2_q$ values change very little all the way to $\beta=0$ ($T\to \infty$) in which limit $\sigma^2_q \to 2/9$ irrespective of $D$. In addition, we observe that $\sigma^2_q$ undergoes a discontinuous, finite jump at the critical point. We also note that the value of $\sigma^2_q$ at the critical boundary on the superfluid side is still small, thus confirming the fact that zero field criticality in the BEG model is powered by the spin-spin correlations only, \cite{beg}. Nonetheless, the discontinuity in $\sigma^2_q$ brings out the effect of criticality on the quadrupole-quadrupole interactions, which appear to be very different in the two phases. The superfluid scalar curvature $R_q$ faithfully reflects the trend in the superfluid $q-q$ interactions near criticlality. In figs.\ref{zfcr2}(b) and (c) we plot {\it along} the critical line, in turn, $\sigma^2_q$ and the superfluid curvature $R_q$ vs. $D$. Both quantities have been plotted at a fixed relative distance $t=(\beta-\beta_c)/\beta=10^{-5}$ within the critical line of fig. \ref{phase1}(a). It is seen that in the vicinity of the critical boundary within the superfluid phase the $q$-curvature $R_q$ and the quadrupole fluctuation $\sigma^2_q$ both remain small, especially for values of $D$ away from the tricritical point at $D_{tcr}=0.462$, thereby underscoring the fact that the correlation length for the order parameter $q$ remains small near criticality so that it does not play any direct role in phase ordering here. Near the tricritical point however, to which we turn now, the case is not the same.

\subsection{Geometry near the tricritical point}
\label{hai2}

\begin{figure}[t!]
\begin{minipage}[b]{0.3\linewidth}
\centering
\includegraphics[width=2in,height=1.3in]{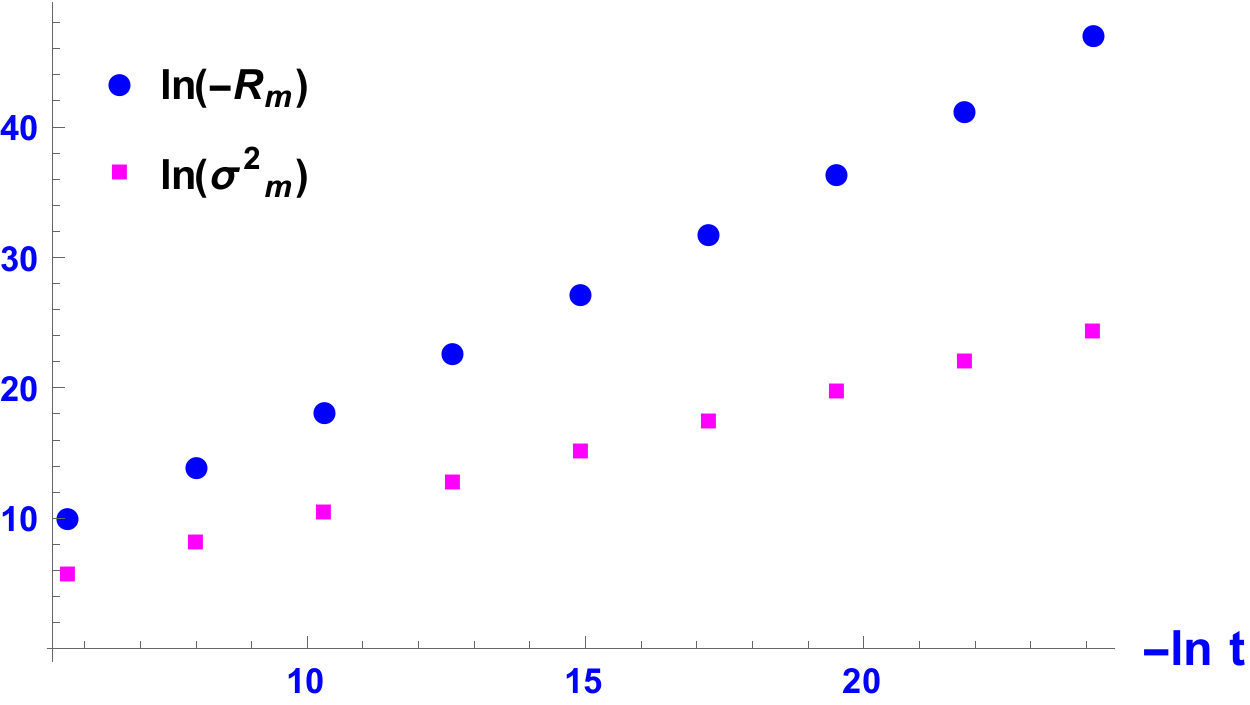}
\end{minipage}
\hspace{0.2cm}
\begin{minipage}[b]{0.3\linewidth}
\centering
\includegraphics[width=2in,height=1.3in]{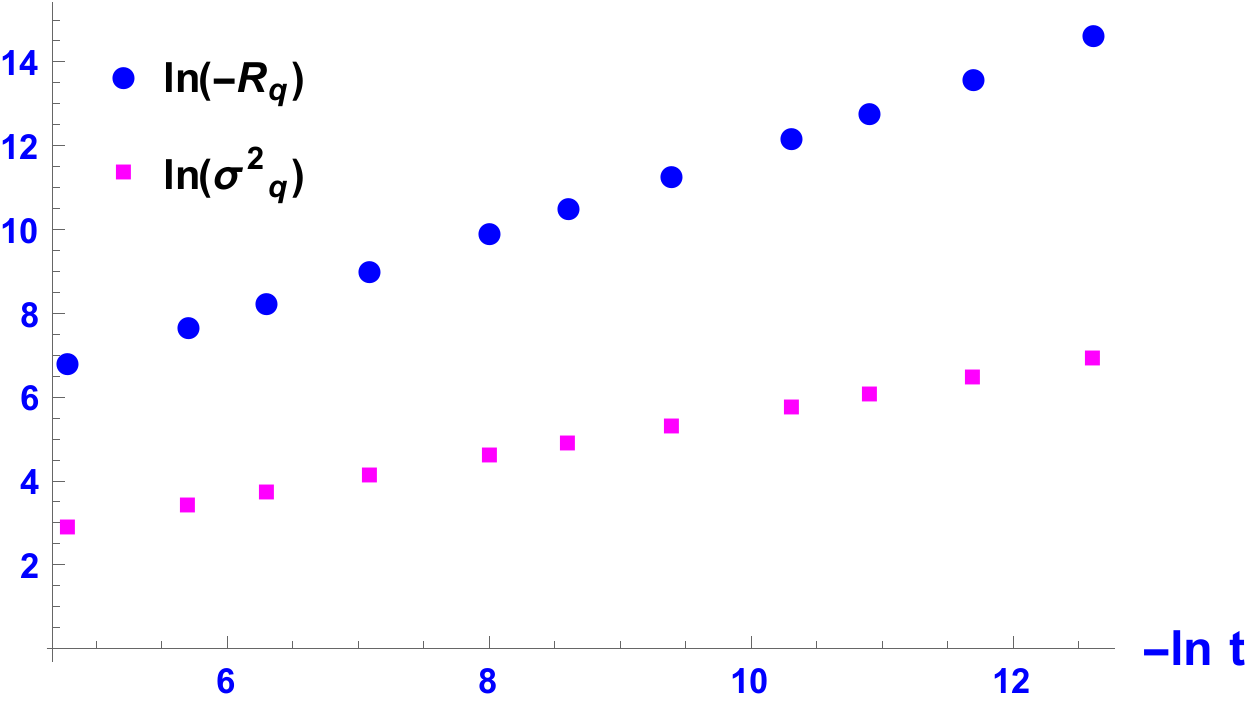}
\end{minipage}
\hspace{0.2cm}
\begin{minipage}[b]{0.3\linewidth}
\centering
\includegraphics[width=2in,height=1.3in]{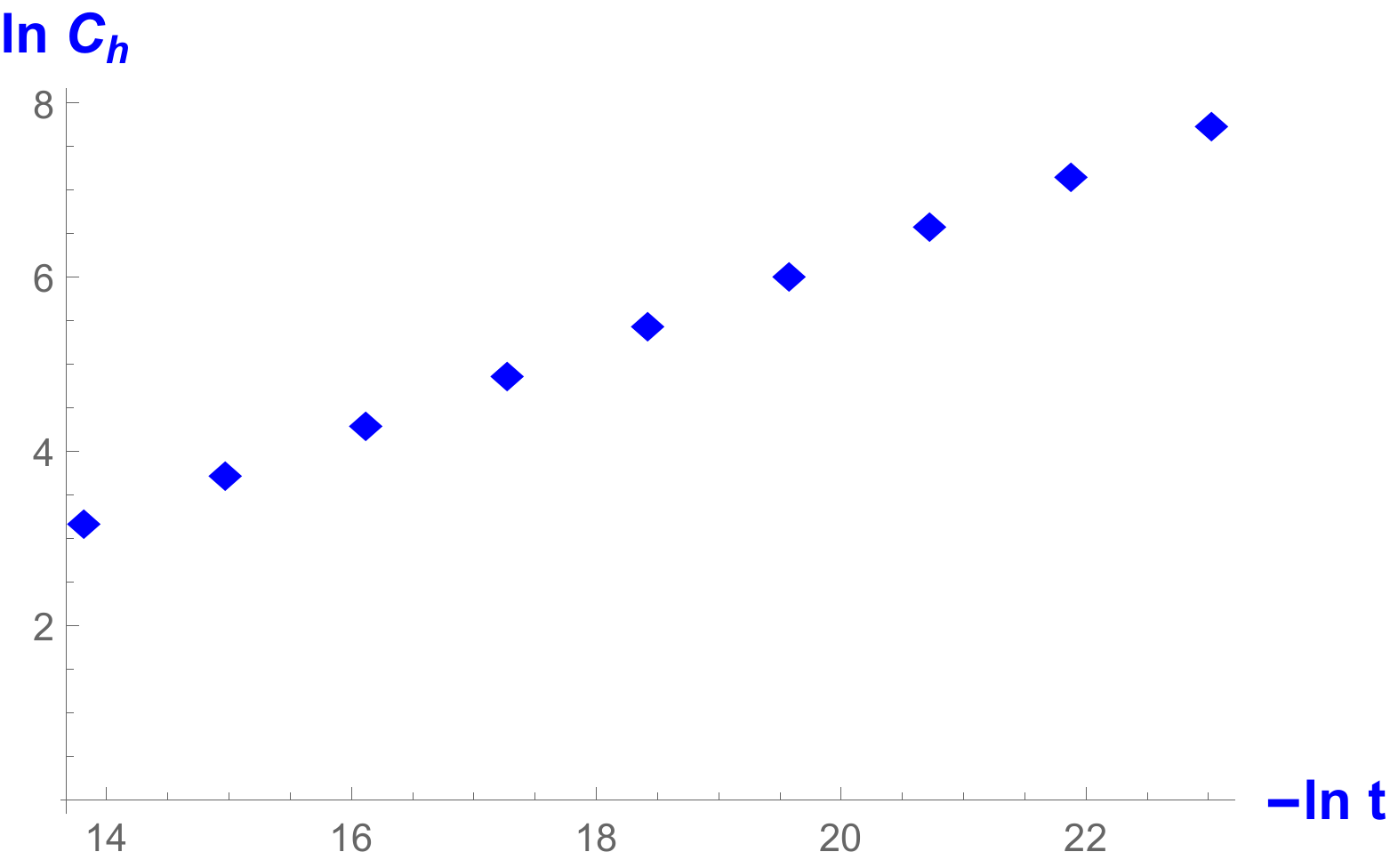}
\end{minipage}
\caption{\small{Sub-figures $(a)$ and $(b)$ depict the scaling behaviour in the superfluid phase near the tricritical point at $D=0.4621,\beta=3$ for the Blume-Capel limit ($\omega=0$). $(a)$ From the log-log plot of $R_m$ and $\sigma_m^2$ vs. $t$ the slope is seen to be $2$ and $1$ respectively. $(b)$ The slope from the log-log plot of $R_q$ and $\sigma_q^2$ vs $t$ is seen to be $1$ and $1/2$. In $(c)$ the slope of the log of the superfluid heat capacity $C_h$ is $1/2$ while in the normal phase $C_h$ remains finite (not shown).  }}
\label{tcr}
\end{figure}

 The tricritical point is the culmination of the triple line so that at this point three phases become one, and in the zero field plane it joins the critical line and the coexistence line, fig.(\ref{phase1}). It is at the junction of three critical lines, the zero field critical line and two symmetric non zero field wing critical lines on the positive and the negative sides of the $H$ axis, fig.(\ref{meanfieldbeg}). The scaling laws at the tricritical point are different from the critical points, even for the mean field case, \cite{chaikin, lawrie}.  For the critical exponent $\beta$, while along the rest of the critical line we have $m\sim t^{1/2}$ on approaching the tricritical point we get $m\sim t^{1/4}$. Similarly, for the critical exponent $\delta$, we have $m\sim H^{\frac{1}{3}}$ for ordinary critical points and $m\sim H^{\frac{1}{5}}$ for the tricritical point as can be quickly ascertained for the Blume Capel case from equations (\ref{critBC}), \ref{tcrp}) and(\ref{magfield}). Interestingly, the scaling exponents at the tricritical point equal their mean field values for dimensions $d=3$ as opposed to the upper critical dimension $d=4$ for ordinary critical points. This can be seen from the formula for the upper critical dimension obtained from the Ginzburg criterion, \cite{lawrie}, $$ d_u=2 (\beta/\nu+1)$$
 The zero field superfluid scalar curvature $R_m$ however continues to diverge at the same rate $$R_m\sim t^{-2}$$ as elsewhere on the critical line so that, in itself, it does not signal anything special happening to the $m-m$ correlations at that point. However, as is apparent from fig.\ref{zfcr2}.(c) the $q$-geometry already senses the vicinity of the tricritical point, with the scalar curvature $R_q$ growing in magnitude for $D$ values close to its tricritical value. This is also true of $\sigma^2_q$ in \ref{zfcr2}.(b). At the tricritical point the superfluid $q$-curvature diverges as $$R_q\sim t^{-1}$$ with the quadratic fluctuation diverging as $\sigma^2_q \sim t^{-1/2}$. We therefore conclude that the $q-q$ correlations become more and more long range as we approach the tricritical point, with $\xi_q$ eventually diverging there, but at a rate different from that of the divergence of $\xi_m$.
We establish the scaling behaviour of the scalar curvatures in the superfluid phase graphically from  fig \ref{tcr}(a) and (b) above. Figure \ref{tcr}(a) presents the log-log plot of $R_m$ and $\sigma^2_m$ vs the reduced inverse temperature $t$. It is easy to establish the scaling behaviour stated above from the log-log plot. Similarly, the log-log plot in fig. \ref{tcr}(b) establishes the scaling behaviour of $R_q$ and $\sigma^2_q$.

 Given that the upper critical dimension for the tricritical point is three one would have expected the scaling of the scalar curvature to be $3/2$ so that in three dimensions it would have led to $\nu=1/2$ as expected. This is also suggested by the scaling of the heat capacity $C_h$ near the tricritical point which goes as $$C_h \sim t^{-1/2}$$ in the superfluid phase as shown in fig. \ref{tcr}(c). Therefore, from the exponent relation $$\nu d=2-\alpha$$
 the scalar curvature ought to scale as $t^{-3/2}$. Curiously, while neither $R_m$ or $R_q$ have this scaling, it is the average of the two that turns out to be $3/2$. While this could well be nothing more than a numerical fluke, we suspect that a three dimensional scalar curvature could have possibly been more apt to probe the tricritical point. We defer further investigation to the future.

 In any case the geometry does inform us that microscopically the tricritical point is different from the rest of the zero field critical line in that here both the $q-q$ and the the $m-m$ correlations play a role in the phase transition, though their respective strengths appear to be different. We again believe that the scaling of $R_q$ too ought to be $3/2$ and hope that a future analysis will shed light on it. The difference in scaling disappears in the wing critical region to which we shall turn in the coming subsection.
 
  Finally, as in eq.(\ref{Rm hscaling}) for the critical case, we obtain the tricritical scaling of $R_m$ in the $D$ plane along the $t=0$ line. It is found to be
\begin{equation}
R_m \sim h^{-8/5}\hspace{0.5in}(t=0, \mbox{ tricritical point})
\label{Rm tricritical hscaling}
\end{equation}
This is different from teh critical value. While the $t=0$ path is uncommon as a measure of scaling behaviour for the correlation length, we do hope that it should be possible to check these with Monte-Carlo simulations or finite size scaling studies. 

\subsection{Geometry near the wing critical line}
\label{hai3}

\begin{figure}[t!]
\begin{minipage}[b]{0.25\linewidth}
\centering
\includegraphics[width=1.9in,height=1.4in]{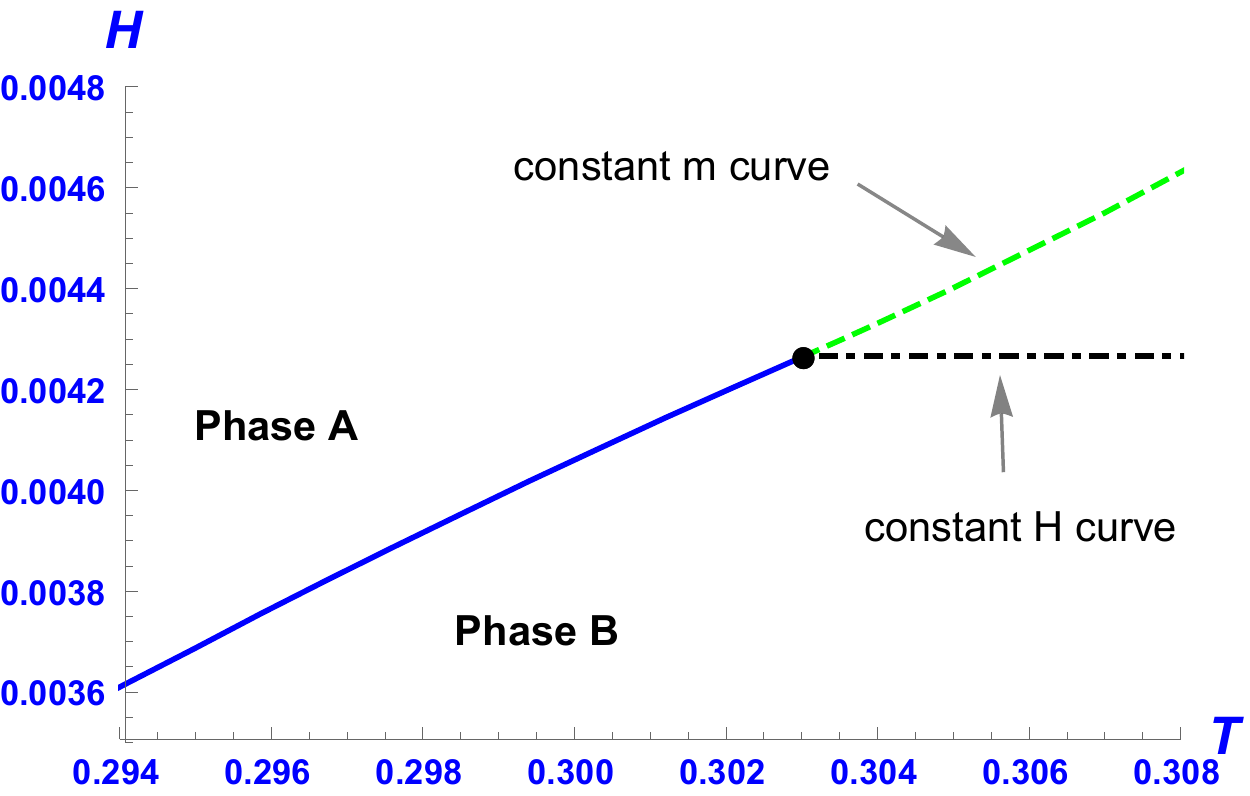}
\end{minipage}
\hspace{1cm}
\begin{minipage}[b]{0.25\linewidth}
\centering
\includegraphics[width=2in,height=1.6in]{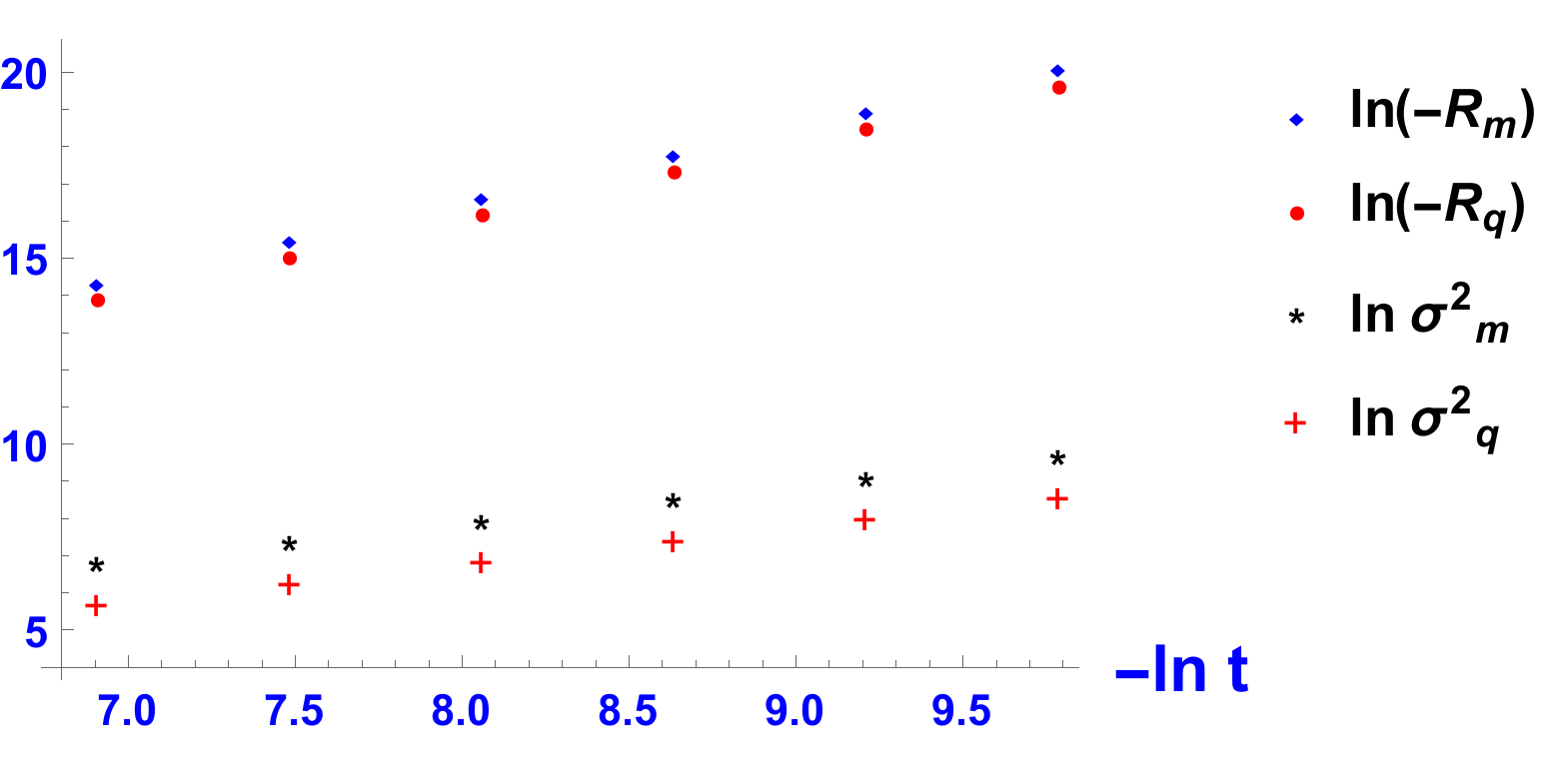}
\end{minipage}
\hspace{1cm}
\begin{minipage}[b]{0.25\linewidth}
\centering
\includegraphics[width=1.95in,height=1.6in]{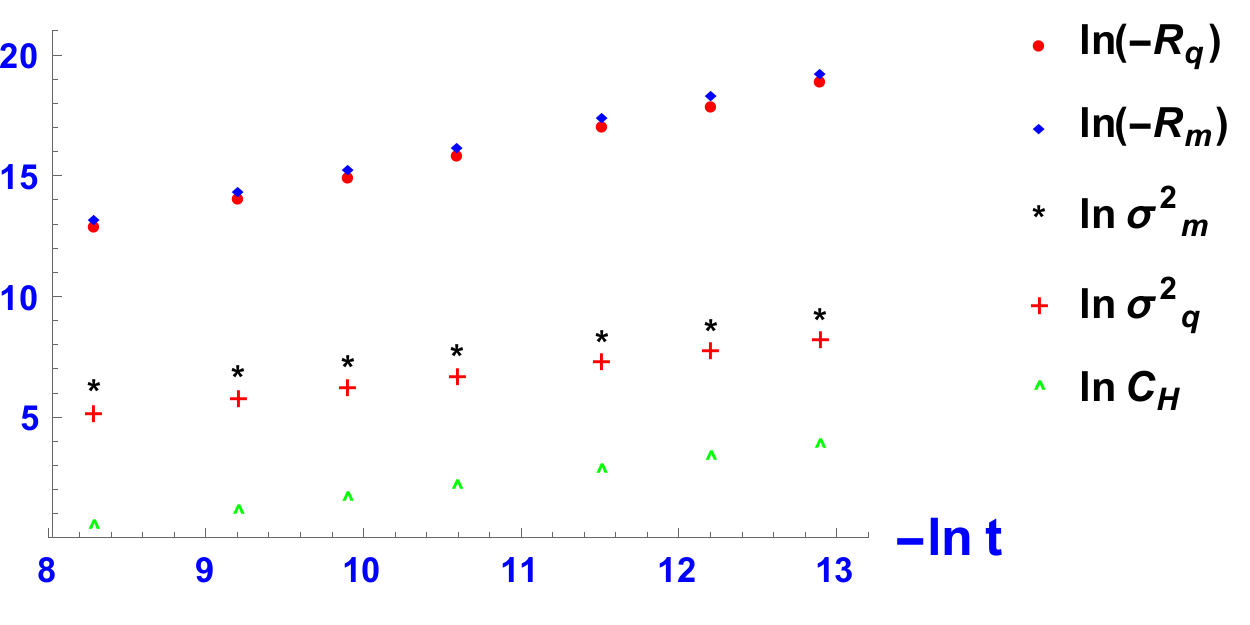}
\end{minipage}
\caption{\small{Subfigure $(a)$ shows the Blume-Capel wing coexistence curve in the $T-H$ plane terminating with the (thick dot) critical point at $D_{cr}=0.474,H_{cr}=0.004,\beta_{cr}=3.3$ and $m_{cr}=0.301$. The coexisting phases are labelled A and B. Two directions of approach from the single phase region to the critical point are shown, one at constant $H=H_{cr}$ and the other at constant $m=m_{cr}$. Log-log plots in $(b)$ and $(c)$ show the scaling behaviour in the $D$ plane along, respectively, the constant $m$ curve and the constant $H$ curve. The slopes of $R_m$ and $R_q$ equal $2$ in $(a)$ and $4/3$ in $(b)$ while those of $\sigma_m^2$ and $\sigma_q^2$ equal $1$ in $(a)$ and $2/3$ in $(b)$. Also shown in sub-figure $(c)$ is a log-log plot of $C_h$ with a slope of $2/3$. }}
\label{wingcr}
\end{figure}

 While the non zero $H$ field is physically unrealistic in the context of Helium mixture it acts as a regular magnetic field in the Blume-Capel model. Here we shall restrict the non zero field investigations to the Blume Capel case. The geometric analysis can easily be extended to non-zero $\omega$ values though we shall not pursue it here. 
 
 The critical line in the non-zero field borders the wing coexistence region as mentioned previously in our description of the phase structure around fig.(\ref{wingphase}). The intersection of the wing surface with the constant $D$ plane (for $D>\log\frac{4}{3}$) appears much like the $T-P$ coexistence line of simple fluids, with the pressure field replaced by the $H$ field and its conjugate density replaced by the magnetic moment. In fig. \ref{wingcr}.(a) we plot a wing coexistence curve in the $T-H$ plane for the Blume-Capel case. It separates two paramagnetic phases labelled `A' and `B' and terminates at a critical point, much like in a van der Waal fluid. We obtain two different scaling behaviours by approaching the critical point from separate directions. In fig. \ref{wingcr}.(b) we obtain log-log plots of $R_m$, $R_q$ and the fluctuations $\sigma^2_m,\sigma^2_q$ along the constant $m$ curve, with the magnetization fixed at its critical value. This is the standard path to ascertain critical exponents from the single phase region, \cite{pathria}. Here we see that $R_m\sim t^{-2}$. Interestingly, $R_q$ too scales in exactly the same manner, thus suggesting that in non-zero field region there is only one correlation length. This is known to be true in the one dimensional spin one model \cite{krinsky} and was also confirmed geometrically in paper I. Meanwhile the heat capacity $C_m$ remains finite near the critical point so that along the constant $m$ path the exponent $\alpha=0$, thus conforming to exponent relation $\nu d=2-\alpha$. We have also checked numerically that curvature $R_m$ and the heat capacity $C_m$ conform to eq. (\ref{rchtsq)})
 \begin{equation}
 R_m\,C_m\,t^2 = -\frac{1}{2} \hspace{0.5in} (\mbox{wing critical point}, m=m_{cr})
 \label{wingcr hyper}
 \end{equation}
which is exactly analogous to the case of the van der Waals fluid, \cite{rupprev}. In fig. \ref{wingcr}.(c) we show the scaling of $R_q, \sigma^2_q$ and $C_h$ along the line $H=H_{cr}$. The heat capacity $C_h$ here is analogous to $C_p$ for fluids. It can be seen that along the constant $H$ line $$R_m \sim t^{-4/3}$$ and $$C_h \sim t^{-2/3}$$ which again conforms to the exponent relation $\nu d=2-\alpha$. The amplitude relation of eq.(\ref{rchtsq)}) is not followed however. We notice from fig. \ref{wingcr}.(c) that once again the curvatures $R_m$ and $R_q$ both have the same scaling.

\section{Geometry of coexistence }
\label{huha}

In addition to encoding critical behaviour thermodynamic geometry is also known to efficiently probe phase coexistence and first order phase transition. This was first discussed in \cite{sahay1} in the context of simple fluids. Following Widom's argument in \cite{widom} equating correlation lengths in coexisting phases near criticality to the interface thickness, it was shown in \cite{sahay1} by extensively using the NIST database for simple fluids that there was an excellent match between the numerically obtained scalar curvatures in the coexisting liquid and vapour phases near the critical point. It was also shown that for a reasonable distance from the critical point the scalar curvature obtained explicitly from the mean field van der Waals model could predict the coexistence curve, thus nicely complementing the Maxwell construction. In the geometric context the coexistence point was obtained by locating the point of intersection of the scalar curvatures and hence (via the Ruppeiner conjecture) the correlation lengths of the two phases. This method of constructing the coexistence curve has hence come to be known as the $R$-crossing method, \cite{may,rcross,ruppmay}. The geometric method advanced in \cite{sahay1} was quickly confirmed by numerical studies based on  equations of state for the Lennard-Jones fluids in \cite{may}. The authors had found striking agreement between the phase envelope obtained by the $R$-crossing method and the one obtained from simulation data. In addition to phase coexistence the geometric investigations have also been fruitful in predicting the Widom line for the supercritical phase, \cite{sahay1,may}, though we will not be pursuing it here. Thermodynamic geometry of phase coexistence has also been investigated in magnetic systems, \cite{sarkar1} and black hole thermodynamic systems, \cite{rishabh,pankajc} among others.

In this section we geometrically investigate the phase coexistence regions in the zero field as well as the across the wings which are situated in non zero magnetic field.

\subsection{Zero field coexistence between the normal and the superfluid phases}
\label{fa}

The first order phase transition in zero field is understood as a phase separation between the normal and superfluid phases in the context of BEG model and between the magnetic and the impurity dominated non magnetic phases in the BC case. In the Blume Capel case and also for small quadrupole-quadrupole coupling ($\omega$ small) in the BEG case the phase separation is governed by superfluid ordering via the order parameter $m$. Thus for $D>D_{tcr}$ the only way to sustain a superfluid order in the Helium mixture (or a magnetic order) is to separate into two phases with He$^3$ atoms (or non-magnetic impurities) dominating the normal phase. For larger values $\omega$ the non magnetic intermolecular interaction force between molecules begins to play an increasingly important role  in phase separation, \cite{beg}. Unfortunately, we shall not be able to examine this trend geometrically since, with the $q$-geometry defined only in the superfluid phase, the $q$-curvature in the normal phase will not be available for comparison. Therefore, we shall be able to study only the $m$-curvature across the zero field phase coexistence.

\begin{figure}[t!]
\begin{minipage}[b]{0.3\linewidth}
\centering
\includegraphics[width=2.6in,height=1.9in]{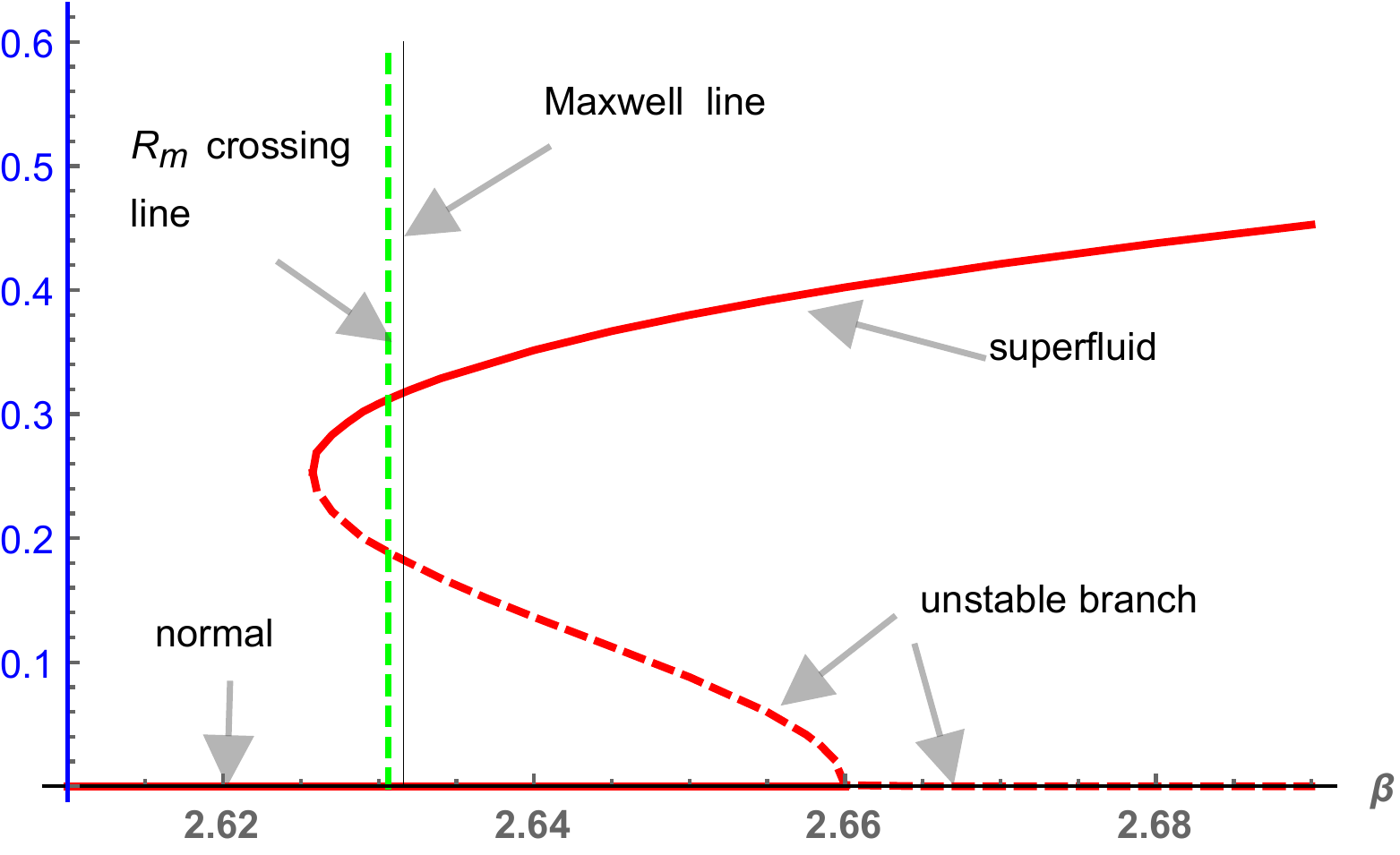}
\end{minipage}
\hspace{3cm}
\begin{minipage}[b]{0.3\linewidth}
\centering
\includegraphics[width=2.6in,height=1.9in]{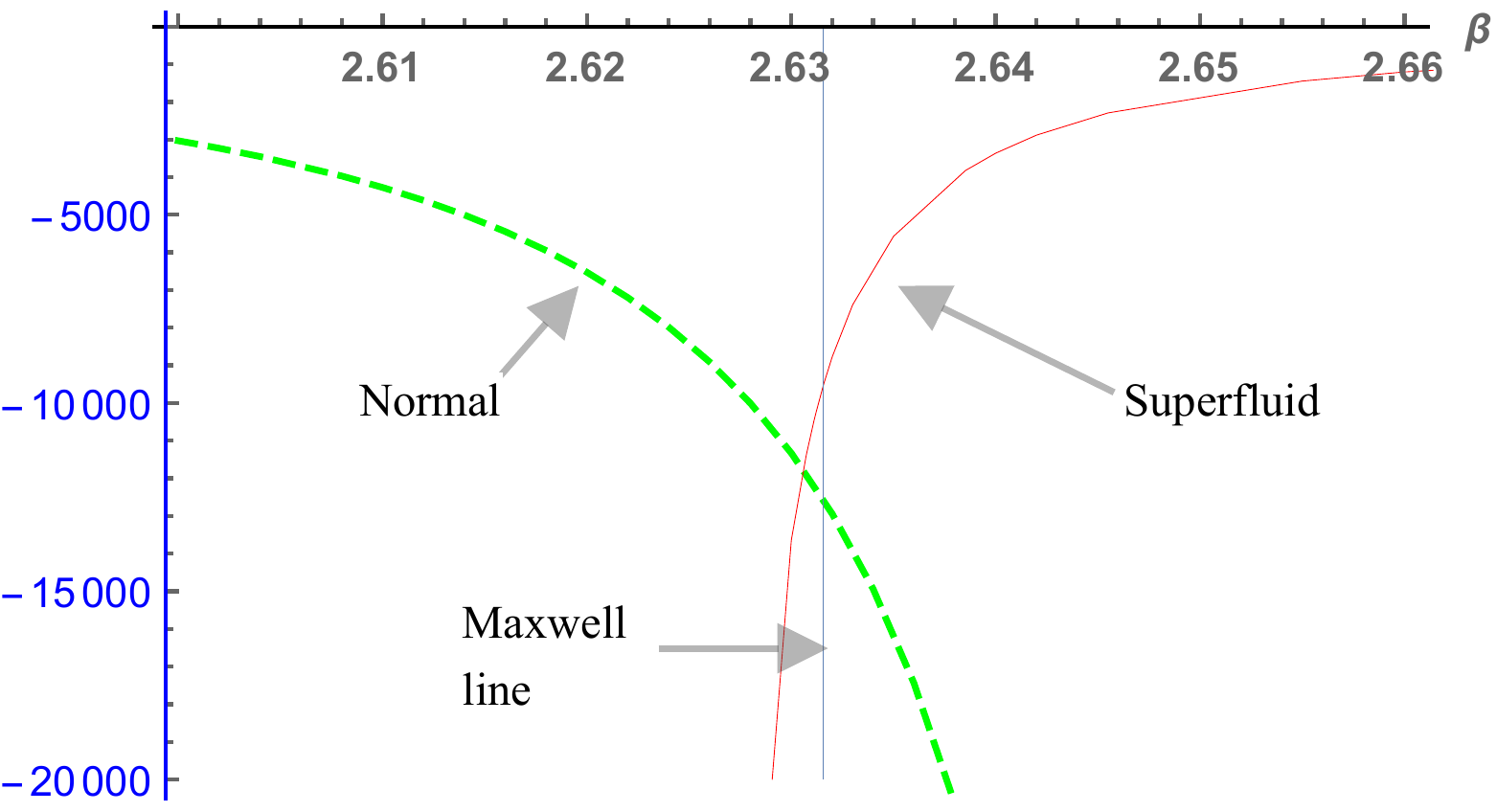}
\end{minipage}
\caption{\small{ Plots comparing $R_m$ crossing to the free energy crossing (referred to as the Maxwell line). Both plots refer to the same first order phase transition in zero field with $D= 0.511$ and at $\omega=0.16$. In $(a)$ the magnetization is plotted vs. the inverse temperature $\beta$. The free energies of the normal and the superfluid phase become equal at the `Maxwell' line $\beta=2.6316$ while the curvatures $R_m$ in the two phases become equal at the $R_m$-crossing line at $\beta=2.6305$. In $(b)$ the curvature $R_m$ for both the phases is plotted and is shown to cross near the Maxwell line at the aforementioned values of $\beta$. At the tricritical point for $\omega=0.16$ the parameter values are $D=0.5044,\,\beta=2.515$.  }}
\label{zfc11}
\end{figure}

In fig.\ref{zfc11}.(a) we plot the magnetization $m$ vs. the inverse temperature $\beta$ in zero field at $D=0.511$ and $\omega=0.16$. For high temperatures (small $\beta$) only the normal phase with $m=0$ exists. On lowering the temperature a metastable superfluid state ($m\neq 0$) begins to coexist with the globally stable (lower free energy) normal phase. On further lowering the temperature beyond the `Maxwell line' the superfluid state achieves global stability with its free energy `crossing' below that of the normal phase which is now metastable. This marks the first order phase transition point as governed by fee energy crossing. On further lowering the temperature the normal phase disappears and the only stable (local as well as global) phase is the superfluid phase. Close to the Maxwell line is a dashed curve labelled the `$R_m$ crossing line' which indicates the temperature at which the $m$-curvatures of the coexisting phases cross each other. This is further elaborated in the curvature $R_m$ vs. $\beta$ plot in fig.\ref{zfc11}.(b), in which the curvatures $R_m$ of the normal and the superfluid phases are seen to cross each other at a temperature close to the free energy crossing temperature indicated by a vertical line.

\begin{figure}[t!]
\begin{minipage}[b]{0.3\linewidth}
\centering
\includegraphics[width=2in,height=1.4in]{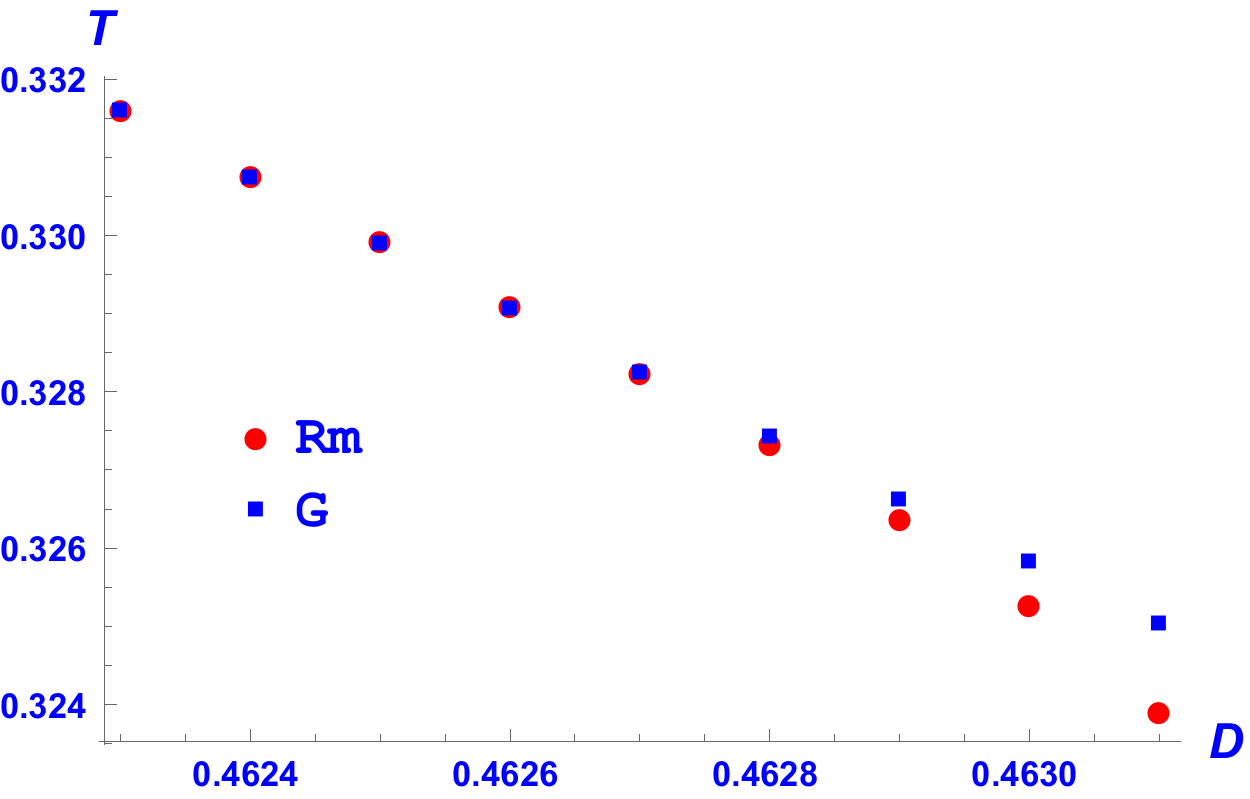}
\end{minipage}
\hspace{0.1cm}
\begin{minipage}[b]{0.3\linewidth}
\centering
\includegraphics[width=2in,height=1.4in]{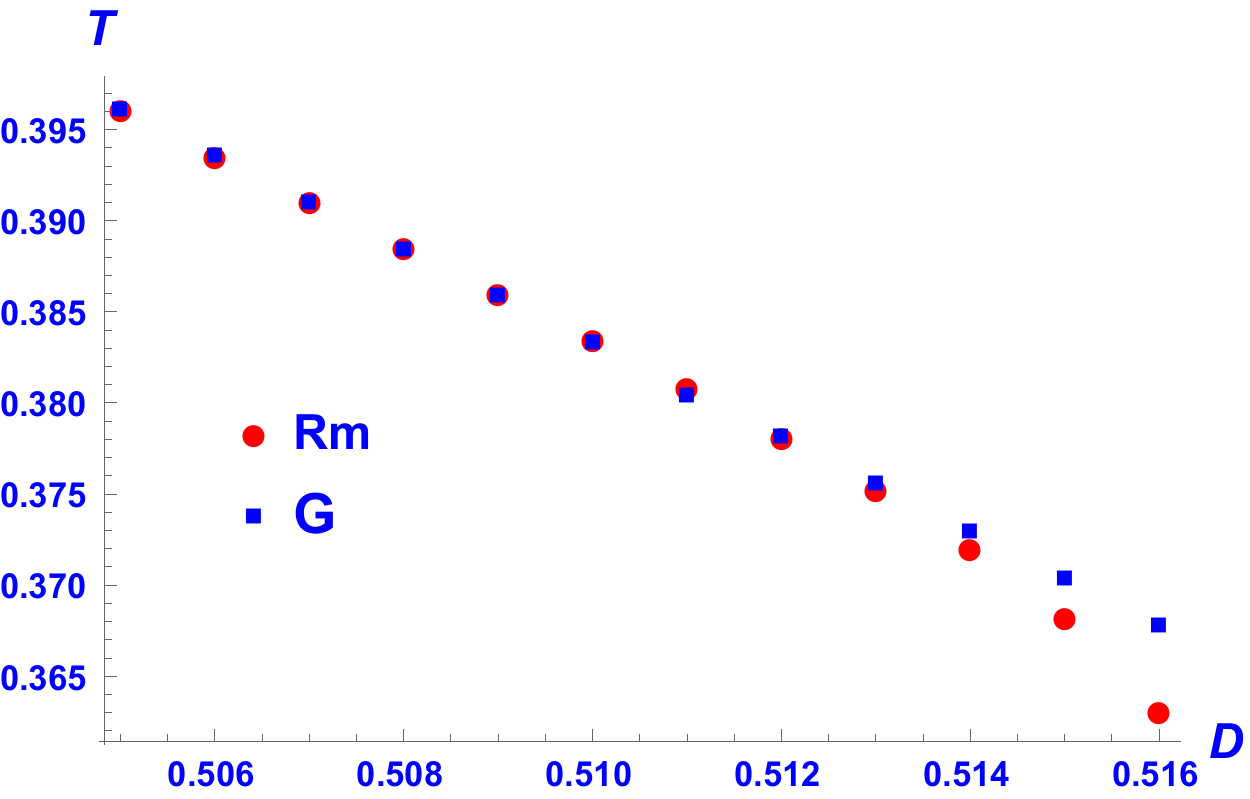}
\end{minipage}
\hspace{0.1cm}
\begin{minipage}[b]{0.3\linewidth}
\centering
\includegraphics[width=2in,height=1.4in]{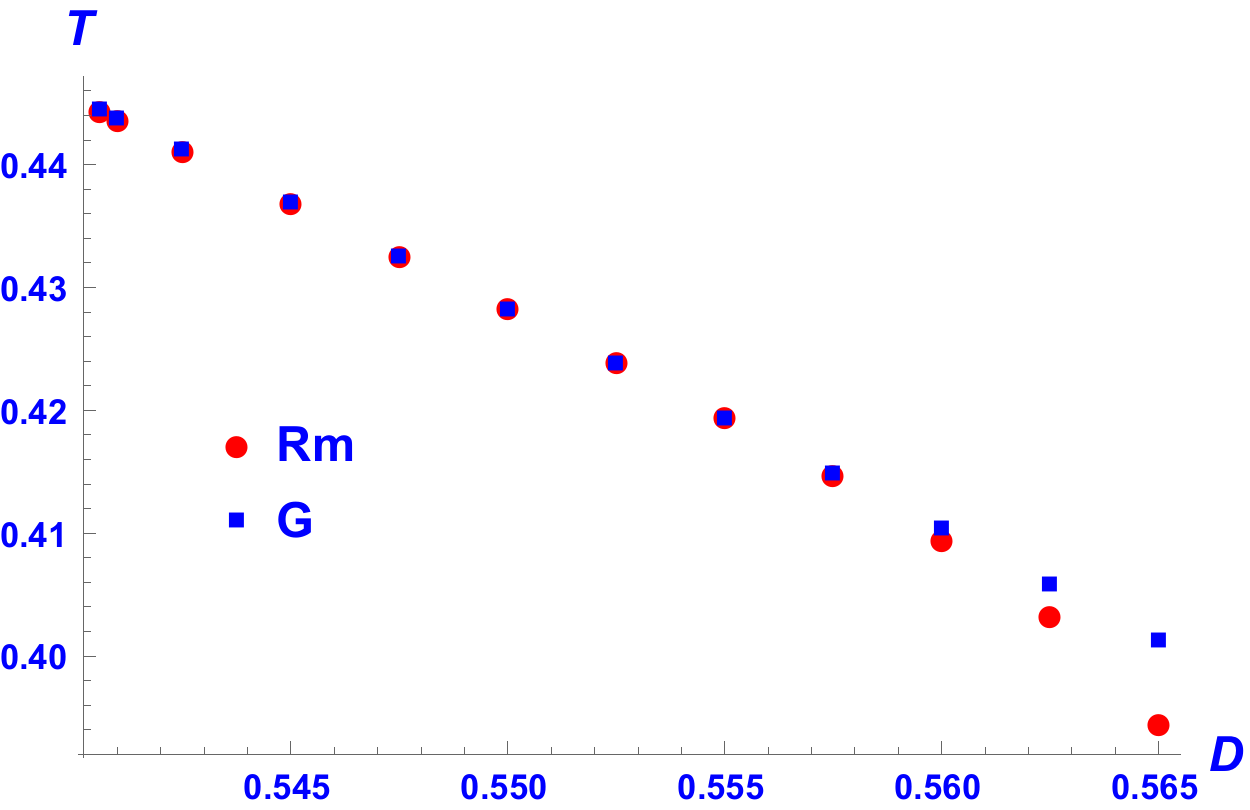}
\end{minipage}
\caption{\small{Free energy crossing points labelled $G$ and the scalar curvature $R_m$ crossing points labelled $R_m$ in the $D-T$ plane for different values of $\omega$. In each case the plots begin at the tricritical point. $(a)$ is the Blume-Capel limit for which $\omega=0$ and the tricritical point is at $D=0.4621,T=1/3$. In $(b)$ $\omega=0.16$ the tricritical point is at $D=0.5044,\,T=0.397$. In $(c)$ $\omega=0.30$ and the tricritical point is at $D=0.5405,\,T=0.445$.  }}
\label{zfc12}
\end{figure}

In each of the sub-figs.\ref{zfc12} (a), (b) and (c) respectively for $\omega=0,0.16,0.3$ we plot in the $D-T$ plane two coexistence curves, the standard one predicted by the free energy crossing, labelled by points `G', and the one predicted by the $R$-crossing method, with points labelled by `$R_m$'. Each of the plots starts from the tricritical point for its $\omega$ value. Remarkably enough, upto a reasonable distance from the tricritical point there is an excellent agreement in the coexistence curves predicted by the free energy and by the curvature crossings. We also notice that as we tune up the quadrupolar interaction by increasing $\omega$ the range of temperatures for which the coexistence curves agree increases. This could probably have to do with the observation in \cite{beg} that increasing the intermolecular interaction strength (quadrupolar coupling in the Hamiltonian) positively induces phase separation which at $\omega=0$ is controlled solely by the spin-spin interactions. Lack of non-trivial $q$-geometry in the normal phase limits our analysis in this case. We hope to pursue this in a future investigation.

\begin{figure}[]
\centering
\includegraphics[width=2.7in,height=1.8in]{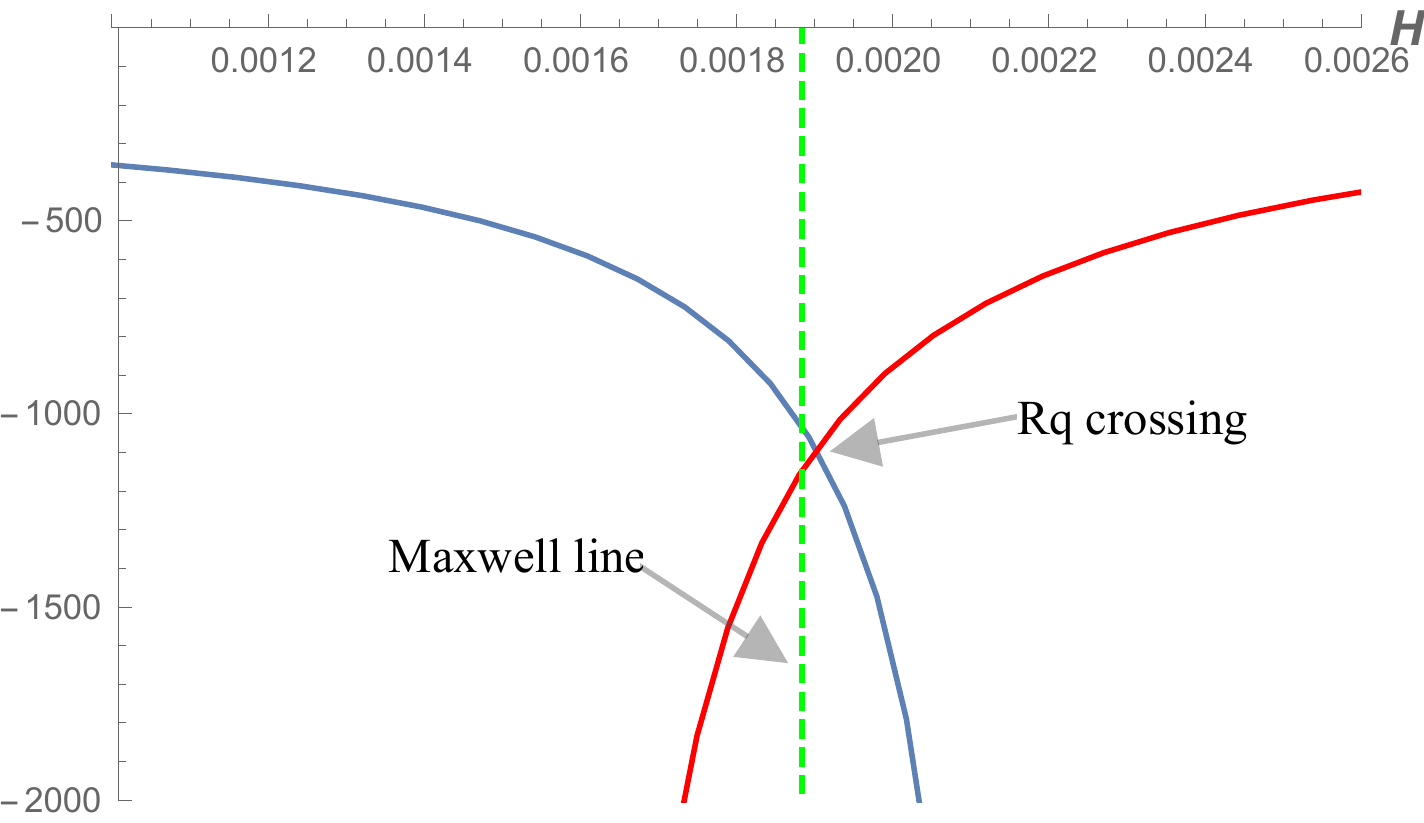}
\caption {{\small $R_q$ crossing diagram across the wing for the Blume-Capel case ($\omega=0$). Plot of $R_q$ vs. $H$ in coexisting phases across the wing region at $D=0.47$ and $\beta=3.3$.}  }
\label{wing1}
\end{figure}

\begin{figure}[t!]
\begin{minipage}[b]{0.3\linewidth}
\centering
\includegraphics[width=2in,height=1.4in]{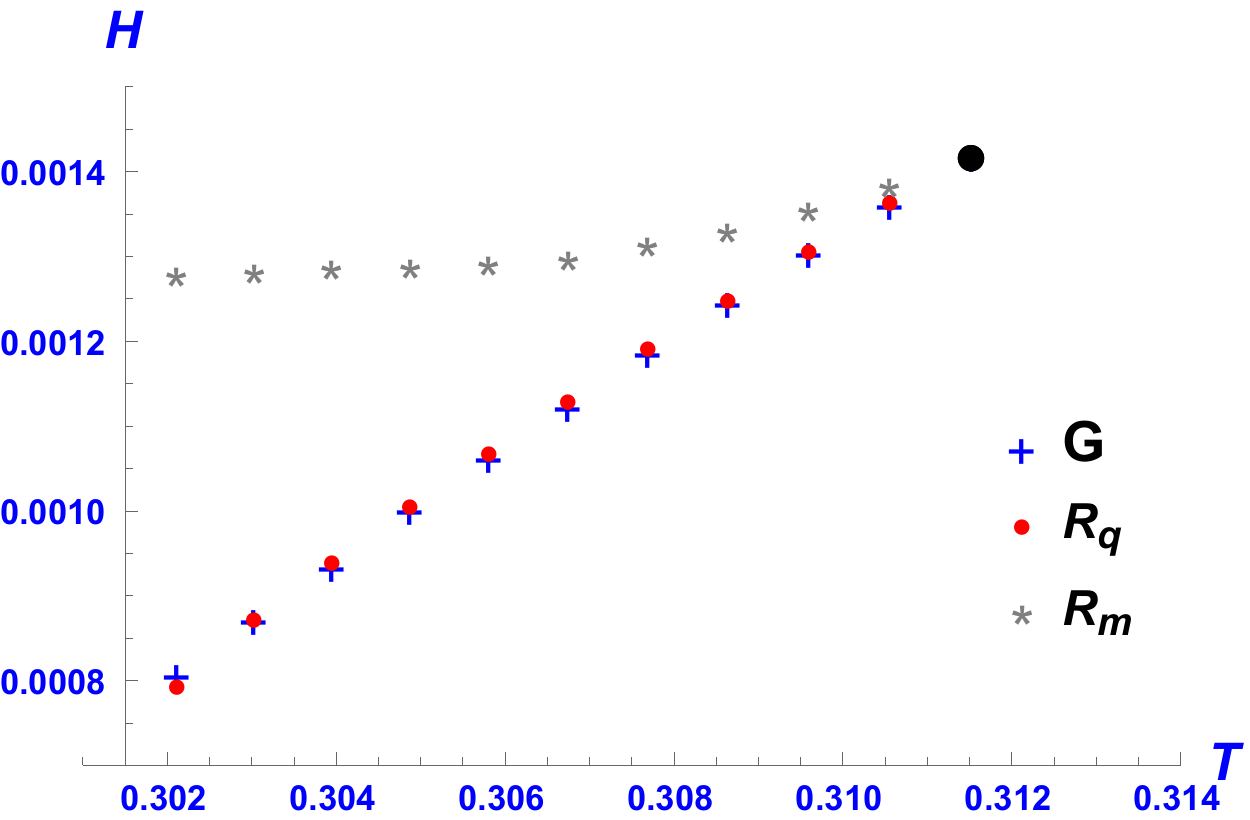}
\end{minipage}
\hspace{0.1cm}
\begin{minipage}[b]{0.3\linewidth}
\centering
\includegraphics[width=2in,height=1.4in]{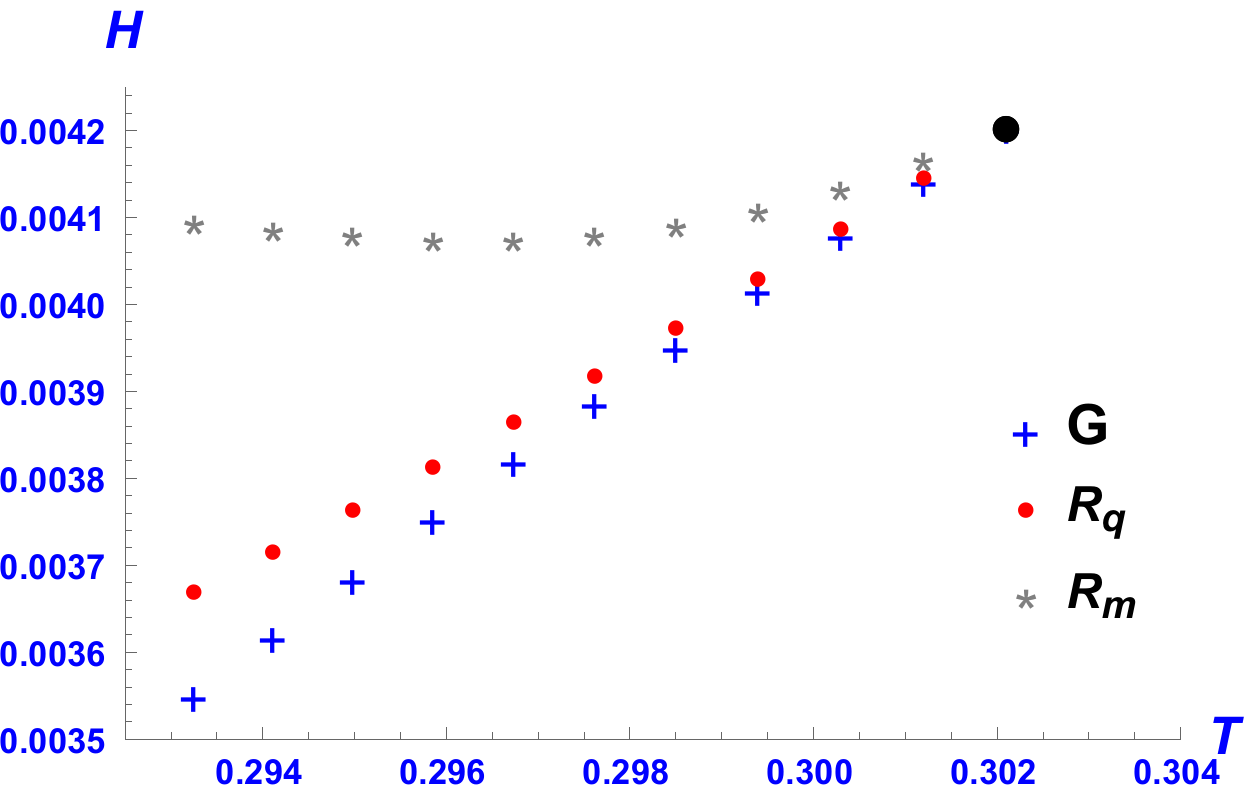}
\end{minipage}
\hspace{0.1cm}
\begin{minipage}[b]{0.3\linewidth}
\centering
\includegraphics[width=2in,height=1.4in]{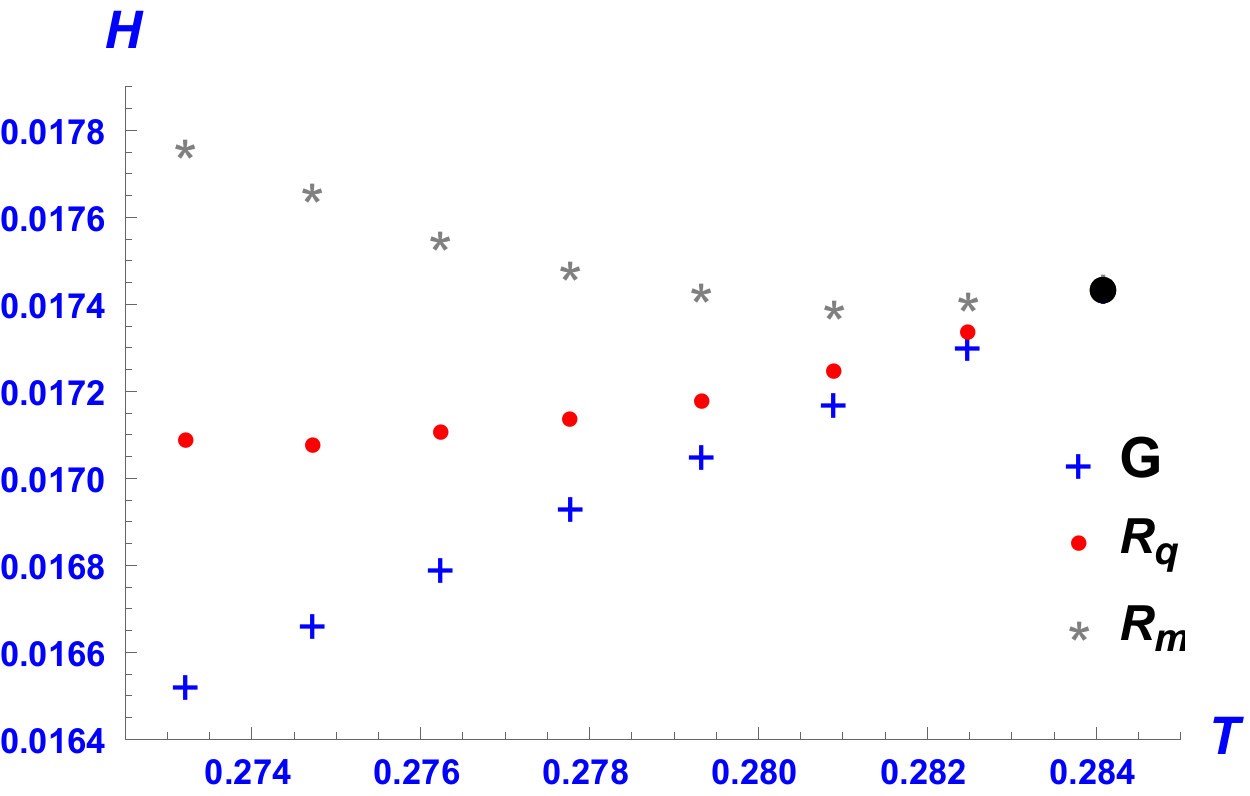}
\end{minipage}
\caption{\small{The free energy crossing points labeled {\it{G}} and the curvature crossing points labeled $R_q$ and $R_m$ in the $H-T$ plane for $(a)$ $D=0.4681$ and $\beta_{critical}=3.2$, $(b)$ $D=0.47413$ and $\beta_{critical}=3.3$ and $(c)$ $D=0.495105$ and $\beta_{critical}=3.5$. The free energy crossing points lie on the coexistence `wing' $B'$ of figure.(\ref{meanfieldbeg}). }}
\label{wing11}
\end{figure}

\begin{figure}[t!]
\begin{minipage}[b]{0.3\linewidth}
\centering
\includegraphics[width=2in,height=1.4in]{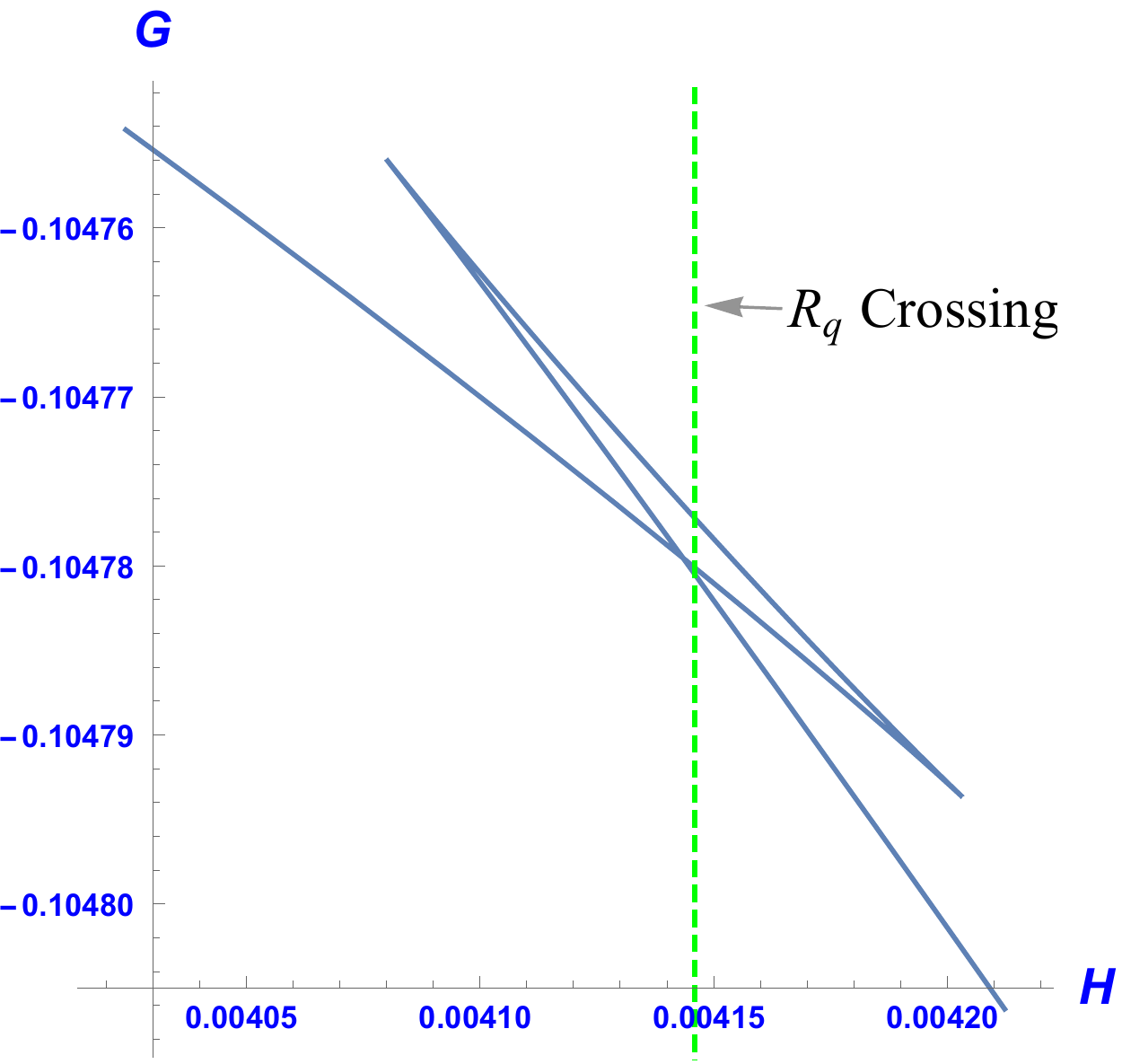}
\end{minipage}
\hspace{0.1cm}
\begin{minipage}[b]{0.3\linewidth}
\centering
\includegraphics[width=2in,height=1.4in]{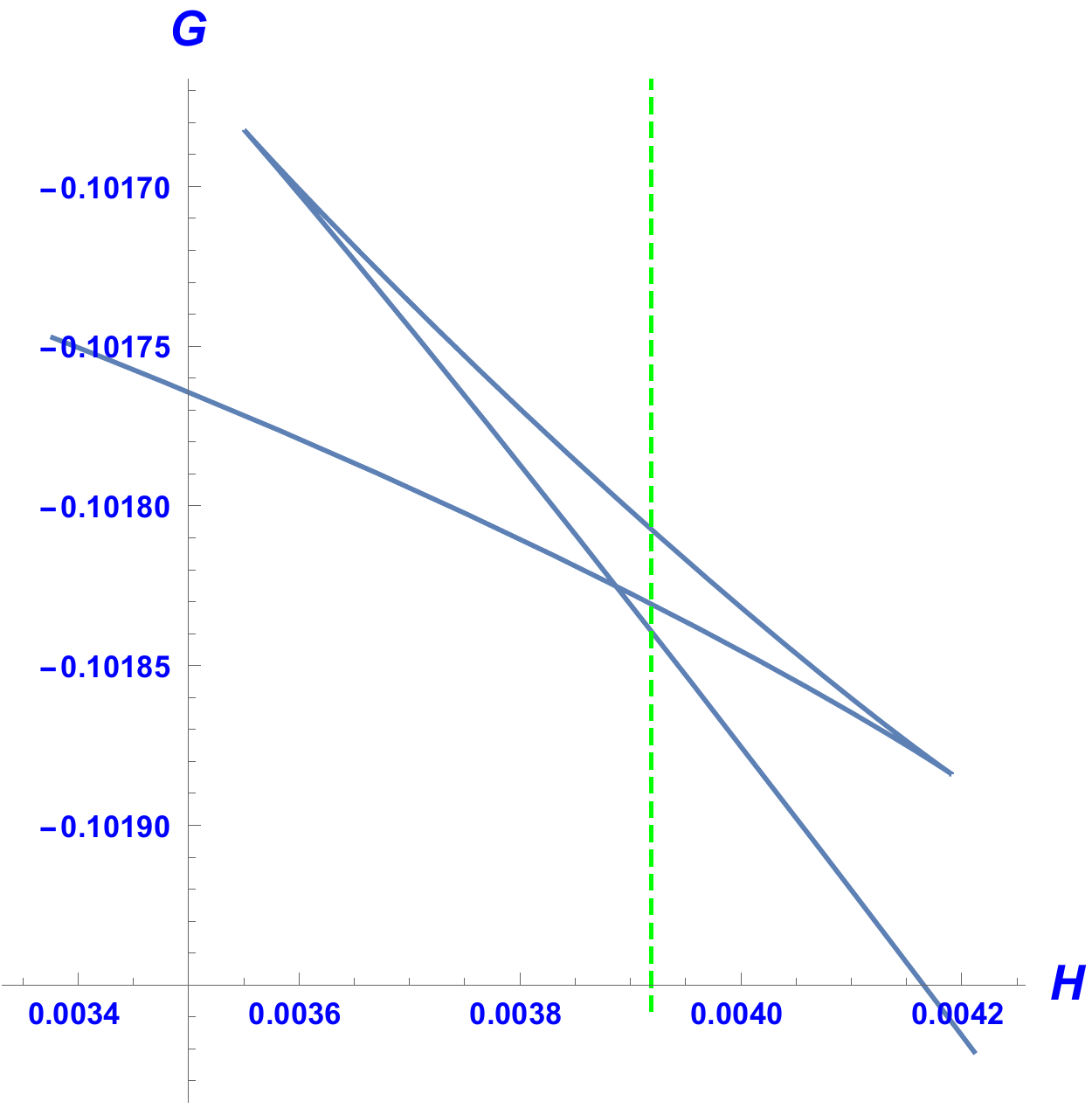}
\end{minipage}
\hspace{0.1cm}
\begin{minipage}[b]{0.3\linewidth}
\centering
\includegraphics[width=2in,height=1.4in]{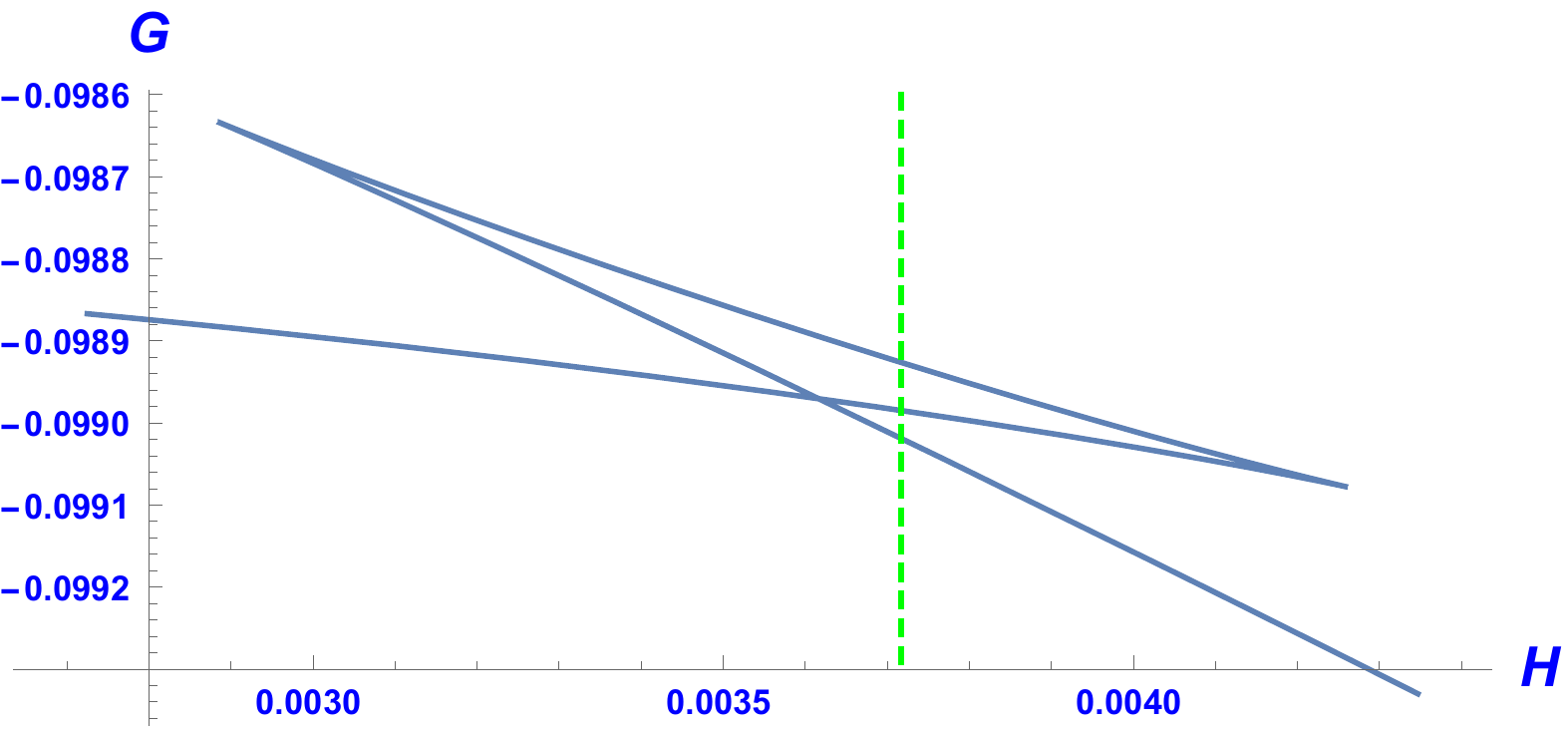}
\end{minipage}
\caption{\small{ Plots of isotherms of free energy in the $G-H$ plane at $D=0.47413$. While the free energy crossing gives the phase transition point via the Maxwell construction, the green dashed line represents the $R_q\,crossing$ point. At the critical point $\beta_{critical}=3.3$. The parameters here are the same as in fig.\ref{wing11}$(b)$. In $(a)$ $\beta=3.32$, the free energy $G$ crosses at $H=0.004144$ while $R_q$ crosses at $H=0.004146$ with a relative separation of $1.62\%, $ in $(b)$ $\beta=3.36$ and the $G$ and $R_q$ crossings are at $H= 0.003888$ and $0.003919$ with separation of $4.84\%$, and in $(c)$ $\beta=3.4$ with $G$ and $R_q$ crossings at $H=0.003619$ and $0.003717$ respectively with a relative separation of $7.11\%$. The values of $R_q$ at the $R_q$-crossing is respectively $-8618,-1345$ and $-644$.}}
\label{wing21}
\end{figure}

\subsection{Coexistence across the wing region}
\label{fu}

 We compare both the $q$-curvature and the $m$-curvature for coexisting phases across the wings. We restrict ourselves to the Blume-Capel case for simplicity though the qualitative picture remains the same for small values of $\omega$.  We shall be plotting the coexistence curves between the two paramagnetic phases (refer to fig.\ref{wingcr}.(a) showing two coexisting paramagnetic phases labelled `A' and `B') in the $T-H$ plane as predicted by the free energy crossing, the $R_m$ crossing and the $R_q$ crossing. Interestingly, it turns out, it is the $q$-geometry whose coexistence curve prediction matches better with the standard coexistence curve obtained via free energy crossing. In fig.\ref{wing1} we plot $R_q$ vs. $H$ for coexisting paramagnetic phases across the wing in the Blume-Capel model. $R_q$ crossing is seen to be close to the free energy crossing. Further, in fig.\ref{wing11} we plot the coexistence curves generated by the self-crossings of the free energy $G$, of $R_m$ and of $R_q$. While the $R_q$ generated coexistence curve is quite close to the standard free energy coexistence curve , the $R_m$ curve seems to diverge quickly. This could probably be an indication that the phase separation across the wing is governed mainly by the quadrupolar interactions rather that the spin-spin interactions. Of course, any such statement must be supported by further analysis, which we defer to a future investigation. In fig.\ref{wing21} we plot the equilibrium  free energy $G$ vs. $H$ for different distances from a wing critical point. The reference figure for these is fig.\ref{wing11}.(b). The free energy self-crossing is compared to the $R_q$ crossing in each sub-figure.  We also note from fig.(\ref{wing11}) that for wing regions closer to the triple line the coincidence between the $R_q$ and the free energy coexistence curves persists for a longer distance from the wing critical point. The discussions around fig.\ref{zfc12} and fig.\ref{wing11} do seem to suggest that the triple line and the tricritical point at its terminus play a strong role in organizing the thermodynamic behaviour around their vicinity. At least the geometry does carry the signature of the triple line in a deep neighbourhood of it.

\section{Conclusions}
\label{ga}

In this work we have extended the geometrical analysis of the one dimensional spin one model initiated in \cite{reiksh} to the mean field BEG model and also its BC limit. We have constructed two complimentary geometries which are demonstrably responsive in turn to the correlations in the magnetization and the quadrupolar order parameters. A detailed scaling analysis near both the zero and the non-zero critical lines concludes that the Ruppeiner equation is followed by the relevant scalar curvatures. The geometry of the tricritical point is also studied and, while the geometry there correctly identifies the role of both the spin and the quadrupole fluctuations, the curvatures there do not scale as expected. Brief comments are made. Geometric coexistence curves are plotted in the zero and non-zero field regions and they are found to agree remarkably well with the standard coexistence curves. In addition, geometry is also able to highlight the difference in the relative importance of one order parameter over the other in ordering the phase dynamics in different regimes.

  We hope that our studies will open the doors to a geometry based analysis of multicritical phenomena and various spin models. Further, it would be good to test and complement the results of our geometrical analysis with Monte Carlo simulations, finite size scaling studies and renormalization group analysis.

\section{Acknowledgement}
\label{gu}

We thank George Ruppeiner for discussions  and his extensive comments on an advanced draft of paper I which also helped refine some of our analysis in this paper. AS thanks Ritu Sharma for discussions and encouragement and DST, Govt. of India for
support through grant no. MTR/2017/001001.

\end{document}